\documentclass[fleqn,usenatbib]{mnras}
\usepackage{newtxtext,newtxmath} 

\usepackage[T1]{fontenc}
\usepackage{ae,aecompl}

\usepackage{graphicx}	
\usepackage{amsmath}	
\usepackage{amssymb}	
\usepackage{makecell}
\usepackage{arydshln}
\usepackage{subfig}
\usepackage{lscape}
\usepackage{multirow}
\usepackage{color}
\usepackage{natbib}

\newcommand\ms{m\,s$^{-1}$}

\newcommand\A{A$_{1}$}
\newcommand\Aa{A$_{2}$}
\newcommand\B{B$_{1}$}
\newcommand\Bb{B$_{2}$}
\newcommand\C{C$_{1}$}
\newcommand\Cc{C$_{2}$}
\newcommand\D{D$_{1}$}
\newcommand\Dd{D$_{2}$}

\defcitealias{grimm2018}{G18}
\defcitealias{delrez2018}{D18}
\defcitealias{haywood2014}{H14}
\defcitealias{Chib2001}{CJ01}

\title[TRAPPIST-1 RV Observations]{Simulating Radial Velocity Observations of Trappist-1 with SPIRou}

\author[B. Klein et al.]{
Baptiste Klein,$^{1,2}$\thanks{E-mail: baptiste.klein@irap.omp.eu}
J.-F. Donati,$^{1,2}$
\\
$^{1}$Universite de Toulouse, UPS-OMP, IRAP, 14 Avenue E. Belin, Toulouse F-31400, France\\
$^{2}$CNRS, IRAP/UMR 5277, Toulouse, 14 Avenue E. Belin, Toulouse F-31400, France\\
}

\date{Accepted. Received; in original form 2019 January 29}

\pubyear{2019}

\begin{document}
\label{firstpage}
\pagerange{\pageref{firstpage}--\pageref{lastpage}}
\maketitle

\begin{abstract}
We simulate a radial velocity (RV) follow-up of the TRAPPIST-1 system, a faithful representative of M dwarfs hosting transiting Earth-sized exoplanets to be observed with SPIRou in the months to come. We generate a RV curve containing the signature of the 7 transiting TRAPPIST-1 planets and a realistic stellar activity curve statistically compatible with the light curve obtained with the K2 mission. We find a $\pm$5~\ms\ stellar activity signal comparable in amplitude with the planet signal. Using various sampling schemes and white noise levels, we create time-series from which we estimate the masses of the 7 planets. We find that the precision on the mass estimates is dominated by (i)~the white noise level for planets c, f and e and (ii)~the stellar actvitity signal for planets b, d and h. In particular, the activity signal completely outshines the RV signatures of planets d and h that remain undetected regardless of the RV curve sampling and level of white noise in the dataset. We find that a RV follow-up of TRAPPIST-1 using SPIRou alone would likely result in an insufficient coverage of the rapidly evolving activity signal of the star, especially with bright-time observations only, making statistical methods such as Gaussian Process Regression hardly capable of firmly detecting planet f and accurately recovering the mass of planet g. In contrast, we show that using bi-site observations with good longitudinal complementary would allow for a more accurate filtering of the stellar activity RV signal.
\end{abstract}





\section{Introduction}

Transiting Earth-sized exoplanets are prime targets to better understand planet formation and evolution. In addition to estimating planet radii from transit depths, one can also measure planet masses through ground-based follow-ups of the host star radial velocity (RV) using high-precision velocimeters. The resulting mass-radius relations are used to constrain the planet interior structures and, beyond that, the whole paradigm of planet formation \citep[e.g.][]{weiss2014,zeng2016,alibert2017,dorn2018}. However, most of Earth-sized planets unveiled with the Kepler space telescope through transit photometry lack mass measurement as they produce low-amplitude RV signatures ($\la$1~\ms) on stars that are too faint for RV surveys in the V band. As a result, the mass-radius diagram of Earth-sized planets is populated by only a handful of planets with well-constrained bulk density \citep{santerne2018}.

M dwarfs are the most promising targets to unveil Earth-like exoplanets with precise masses and radii. Beyond the fact that they largely outnumber stars with earlier spectral type in the solar neighborhood \citep{henry2006}, they feature smaller sizes and masses, and more compact habitable zones (HZ), making HZ terrestrial planets easier to detect as well as rosy candidates for further atmosphere characterization with the JWST \citep{morley2017}. Over the past few years, a growing number of attractive Earth-sized exoplanets have been discovered around M dwarfs from photometric surveys \citep{bertathompson2015,gillon2017,dittmann2017}. This trend is expected to step up with the ongoing TESS mission \citep{ricker2015} as M~dwarfs are known to frequently host multiple terrestrial planetary systems \citep[e.g.][]{bonfils2013,dressing2015,gaidos2016}.

The Spectro-Polarimetre Infra-Rouge \citep[SPIRou; see][for a complete review of the instrument and related science]{donati2018} is a near infrared (nIR) \'echelle spectropolarimeter at the Canada France Hawaii Telescope (CFHT) whose science observations have recently started. The combination of a resolving power of $\sim$70~000 over the YJHK bands and a goal RV precision of $\sim$1~\ms\ makes SPIRou ideally suited for detecting Earth-twins around M dwarfs, not least in the HZ where they typically produce a RV stellar reflex motion of $\sim$1~\ms\ \citep{artigau2018}. 

The SPIRou Legacy Survey (SLS) is a 300n programme dedicated to the main science goals of SPIRou, with 75n focussing on the RV follow-up of transiting planets around M dwarfs. This component of the SLS (called the SLS-TF for SLS transit follow-up) is expected to provide accurate mass measurements for the most attractive transiting planets to be unveiled by a wide range of photometric surveys (e.g. TESS; K2, \citet{howell2014}; MEarth, \citet{nutzman2008}; ExTrA, \citet{bonfils2015}; TRAPPIST, \citet{gillon2017}; NGTS, \citet{wheatley2018}). SPIP\footnote{\url{http://www.tbl.omp.eu/INSTRUMENTATION2/spip}}, the upcoming SPIRou's twin at Telescope Bernard Lyot (TBL at Pic du Midi observatory), will provide additional means to refine the observation strategies of such follow-ups.

Transit photometry provides a priori information on the orbits of the planets around host stars to be monitored within the SLS-TF that can be used to optimize the RV follow-up strategy. Various strategies of RV curve sampling have already been studied in the literature \citep[e.g.][]{ford2008,burt2018}. It turns out that uniform samplings of planet RV curves perform better than more specific strategies, e.g. aiming at orbital quadratures where RVs are expected to be largest.

Besides, a large fraction of M dwarfs exhibits strong magnetic activity resulting in significant photometric and spectroscopic fluctuations \citep[e.g.][]{basri2010,west2011}. The induced RV signal tends to mimic planet signatures, making the former extremely troublesome to filter \citep[e.g.][]{saar1997,queloz2001,Desort2007,dumusque2011,borgniet2015}. If not properly modeled and filtered out, this stellar activity signal is expected to strongly degrade the RV characterization of the SLS-TF targets. Moreover, it imposes additional constraints on the temporal sampling of the RV follow-up, that also needs to densely cover both the rotational cycle of the host star as well as the typical timescale on which it evolves. Last but not least, it makes it extremely hard to recover planets whose orbital periods lie close to the stellar rotation period or its harmonics.



In this paper, we simulate a RV follow-up of TRAPPIST-1 (2MASS J23062928-0502285), an ultracool M dwarf hosting seven transiting terrestrial planets including 3 in the HZ \citep{gillon2016,gillon2017} and possibly more non-transiting ones \citep{kipping2018}. Both the star and the planetary system are well-constrained from photometric surveys \citep[][hereafter \citetalias{grimm2018}]{luger2017,vangrootel2018,delrez2018,grimm2018} and the stellar light curve exhibits quasi-periodic fluctuations indicating the presence of active regions at the surface of the star \citep{vida2017} which makes this system a representative of the SLS-TF targets. Moreover, TRAPPIST-1 will be observed as part of SLS-TF targets in order to provide Transit Timing Variations (TTV)-independent planet mass estimates, still highly-required given the significant differences between TTV and RV planet mass measurements reported in the recent literature \citep[see][and references therein]{mills2017}.


We describe in Section \ref{sec:section2} the simulation of a SPIRou RV follow-up of TRAPPIST-1 involving two distinct steps. First, we synthesize a TRAPPIST-1 RV curve containing both the planet signatures and a realistic stellar activity component scaled on the \textit{K2} lightcurve \citep{luger2017}. We then select the data points according to various sampling schemes to build a range of datasets with different properties. In Section \ref{sec:section3}, we detail our model to recover the planet masses from a given dataset. We present our results for the different sampling configurations in Section \ref{sec:section4} and finally discuss the results and conclude in Section \ref{sec:section5}.

\section{Synthetic datasets}\label{sec:section2}

We first synthesize a densely-sampled RV curve containing both the planetary signature and a realistic realization of the stellar activity RV signal. We then use various sampling schemes to select observational datasets to which we add random values in order to account for instrument and photon noises. The main stellar parameters used in this paper are indicated in Table \ref{tab:star}.

\begin{table}
    \centering
    \begin{tabular}{c|c}
        \hline
        Quantity & Value \\
        \hline
        $M_{*}/M_{\sun}$ & 0.089 \\
        $R_{*}/R_{\sun}$ & 0.121 \\
        $T_{\mathrm{eff}}$ (K) & 2516 \\
        \hline
    \end{tabular}
    \caption{Best estimates of TRAPPIST-1 mass, radius and effective temperature from \citet{vangrootel2018}.}
    \label{tab:star}
\end{table}

\subsection{Planetary RV signature}

Based on the low planet eccentricities reported in \citetalias{grimm2018} and the planetary system stability discussed therein, we consider the simple case of seven planets with circular coplanar orbits assuming no planet-planet interaction. The orbital periods are set to the best estimates from \citet[][hereafter \citetalias{delrez2018}]{delrez2018}. We used Kepler's third law to derive the RV semi-amplitudes from the stellar mass measured by \citet[][see Table \ref{tab:star}]{vangrootel2018} and the last estimates of the planet masses from TTVs \citepalias[see][]{grimm2018}. Initial phases were injected to the planetary orbits using the mid-transit times from \citetalias{delrez2018} with respect to their time reference. The orbital period, mass, semi-amplitude and initial phase of the injected planets are shown in Table \ref{tab:planets}.

\begin{table*}
\centering
\caption{\label{tab:planets} Key parameters of the injected RV signatures of the seven TRAPPIST-1 planets listed in a decreasing order of RV semi-amplitude. The orbital periods are taken from \citetalias{delrez2018}. The planetary masses are the best estimates from \citetalias{grimm2018} and the RV semi-amplitudes are derived using Kepler's third law. The orbital phases are computed from the mid-transit times reported in \citetalias{delrez2018} (T$_{0}$~-~ 2,450,000~[BJD$_{\mathrm{TDB}}$]) assuming circular orbits. The bottom row shows the planet orbital phase shifts roughly corresponding to the maximum TTV predicted by \citetalias{grimm2018} models (see Fig~2 therein).}
\resizebox{\textwidth}{!}{
\begin{tabular}{ll|lllllll}
\hline
\textbf{Planet} & & b & c & g & f & e & d & h \\
\hline
\textbf{Orbital period} & [d] & 1.51087637 & 2.42180746 & 12.354473 & 9.205585 & 6.099043 & 4.049959 & 18.767953 \\
\textbf{Mass} & [M$_{\earth}$] & 1.017 & 1.156 & 1.148 & 0.934 & 0.772 & 0.297 & 0.331\\
\textbf{RV semi-amplitude} & [\ms] & 2.838 & 2.757 & 1.590 & 1.427 & 1.353 & 0.597 & 0.399\\
\textbf{Initial phase} & [rad] & -1.80 & 0.45 & -1.27 & -0.57 & 1.60 & 2.32 & -0.18\\ 
\textbf{Maximum phase-shift predicted from \citetalias{grimm2018}} & [rad] & 0.006 & 0.007 & 0.018 & 0.028 & 0.021 & 0.011 & 0.006 \\ 
\hline
\end{tabular}}
\end{table*}

\subsection{Synthesis of the stellar activity RV signal}

TRAPPIST-1 light curve obtained as part of campaign 12 of the NASA \textit{K2} mission exhibits quasi-periodic fluctuations of $\sim$10~mmag peak-to-peak \citep{luger2017}, symptomatic of the presence of active regions at the stellar surface \citep{vida2017}. Such stellar activity is expected to produce a significant RV signal that needs to be taken into account in a realistic way when simulating the RV follow-up of the star. By synthesizing simultaneously a photometric curve and the corresponding RV curve, we ensure that the dataset we produce is broadly consistent with the statistical properties of the K2 light curve of Trappist-1. This implies that we neglect every RV perturbation induced by the convective blueshift which is anyway expected to be strongly damped for M dwarfs compared to Sun-like stars \citep[e.g.][]{meunier2017}. This also implies that we ignore the Zeeman splitting when computing the line profile, as the magnetic field of TRAPPIST-1 remains poorly-constrained.

We model the stellar surface by a dense grid of 20~000 cells with identical projected area when crossing the meridian. With the aim of including dark and bright inhomogeneities at the stellar surface, we assign a relative brightness factor to each of the cells. For each grid cell, we compute the local intensity line profile using the Unno-Rachkovski analytical solution of the radiative transfer equation, assuming a plane-parallel Milne-Eddington stellar atmosphere \citep{unno1956}. Each local line features a continuum level that is weighted by limb darkening; moreover, it is Doppler-shifted to the projected rotational velocity. The summation of all the local line profiles from the different cells weighted by the local brightness factors provides a global line profile that may be regarded as the cross-correlated function (CCF) of the spectrum. We then derive the RV at the corresponding phase by fitting a Gaussian to the global profile. The relative photon flux is computed by summing the limb-darkened brightness factors of all cells. Our approach for synthesising the stellar activity RV signal slightly differs from the one used in \citet{boisse2012} in that we can more easily include multi-scale magnetic fields and differential rotation in the stellar model, as well as the Zeeman broadening and polarized Zeeman signatures of the modelled line profiles, found to be reliable proxies for diagnosing and modeling stellar magnetic activity \citep{hebrard2016,haywood2016}.

\begin{figure}
\includegraphics[width=\columnwidth]{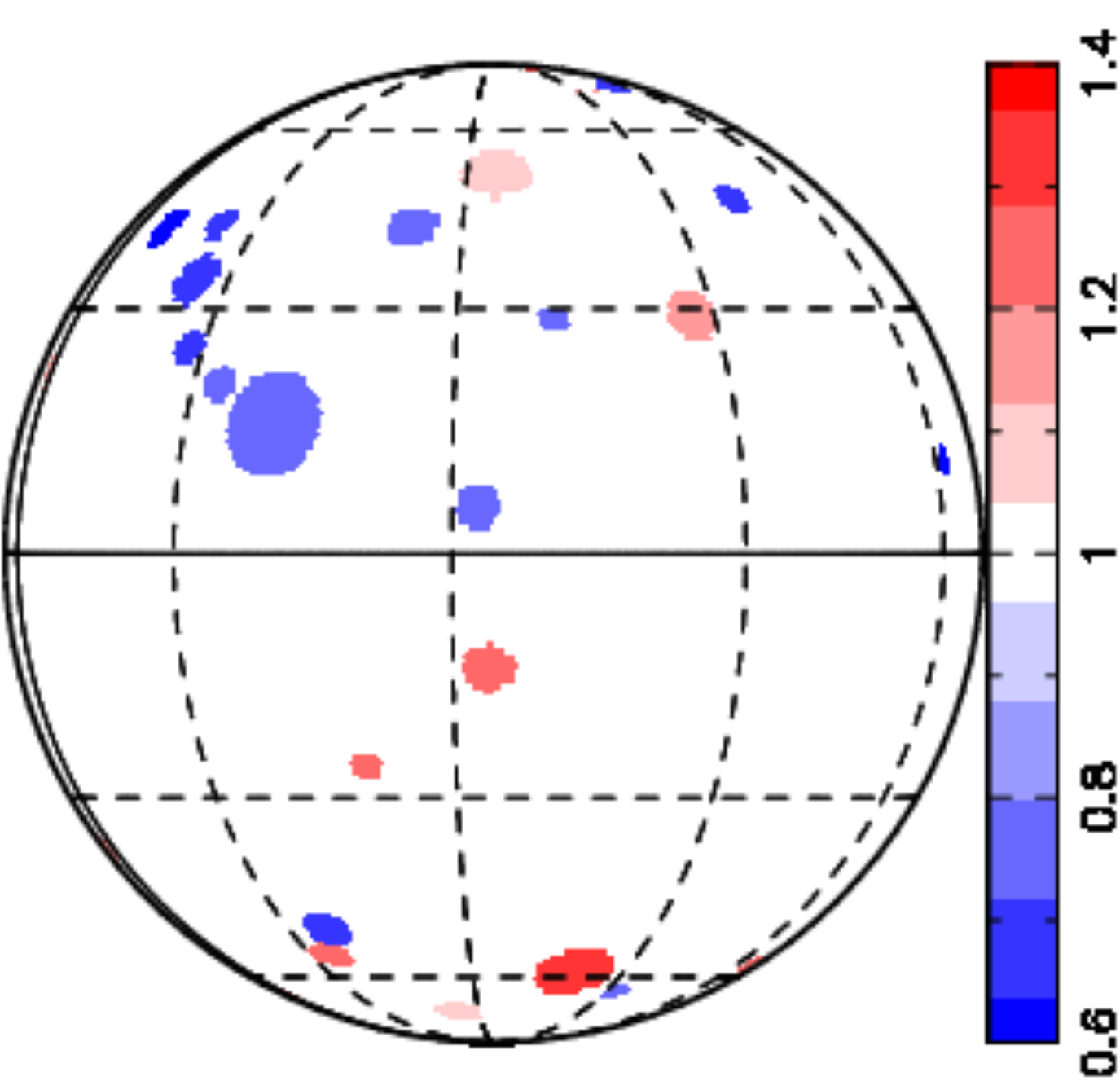}
\caption{Illustration of the modeled stellar surface at a given time. The color of the inhomogeneities indicates their relative brightness: 1 for the quiet photosphere (white), >1 for a bright feature (red) and < 1 for a dark feature (blue).}
\label{fig:star_surface}
\end{figure}

The synthetic star is assumed to rotate as a solid-body with a period of 3.3 days \citep{luger2017}. The time step for the generation of the RV and photometric data points is set at 0.01~d. Our model allows for a dynamical evolution of the active regions at the stellar surface. At each time step, an active region is added to the model with a given probability. The position of the feature is randomly chosen at the stellar surface as active regions are likely to be found at all latitudes on late M dwarfs \citep{barnes2001a,barnes2001} and, in particular, on TRAPPIST-1 \citep{delrez2018,morris2018b,ducrot2018}. Consistently with \citet{morris2018a}, each newly-created feature is randomly chosen to be either brighter or darker than the stellar photosphere. Its relative brightness factor with respect to the stellar photosphere is scaled from \citet[][$\Delta$T~=~$\pm$100 K]{berdyugina2005} using Plank's law in the K2 spectral band. The injected inhomogeneity is circular on the line of sight with a relative size $S_{\mathrm{r}}$ with respect to the whole stellar surface. The radius of each new active region is initially set to zero, increases linearly during the first third of its lifetime and decreases linearly during its remaining lifetime. After that, the feature is removed from the star. The radius of the feature at the time of its maximum size is drawn from a normal distribution centered at 0.1\% to ensure small active regions on the stellar surface in agreement with \citet{rackham2018} and \citet{morris2018b}. Surface features are allowed to overlap each other. As a result, our modeled star is populated by a patchwork of dark and bright small features as illustrated in Fig \ref{fig:star_surface} at any given time.




We simultaneously synthesized stellar photometric and RV curves over a 400-d period using the aforementioned time step. We removed a 100-d burn-in period to ensure that the surface covered by active regions is more or less stable in time with the appearance of new feature roughly compensating the disappearance of old ones. We tuned both the lifetime and probability of appearance of the injected features so the simulated light curve scales with that obtained from the \textit{K2} mission. The main attributes of the injected features on the stellar surface are listed in Table \ref{tab:spotPatterns}.

\begin{table}
\centering
\caption{\label{tab:spotPatterns}Parameters of the features generated at the stellar surface for synthesizing stellar photometric and RV curves. The Jeffreys's distribution is defined in \citet{gregory2007}.}
\begin{tabular}{ccc}
\hline
\textbf{Parameter} &  \textbf{Value} & \textbf{Unit} \\
\hline
\makecell{Probability of new feature\\at each time step} &  0.04 & -- \\
Feature lifetime & $\mathcal{N}(5.0,0.5) \text{ } $ & [P$_{\text{rot}}$]\\
\makecell{Relative brightness dark spot} & Jeffreys(0.6,0.8) & --\\
\makecell{Relative brightness bright spot} & Jeffreys(1.1,1.3) & --\\
\makecell{Latitude of the feature} & $\mathcal{U}(-90,90)$ & [deg]\\
\makecell{log$_{10}$ maximum relative size} &  $\mathcal{N}(-3.0,0.3)$ & --\\
\makecell{Model for spot size evolution} & Linear & --\\
\hline
\end{tabular}
\end{table}

\begin{figure*}
	\includegraphics[width=\linewidth]{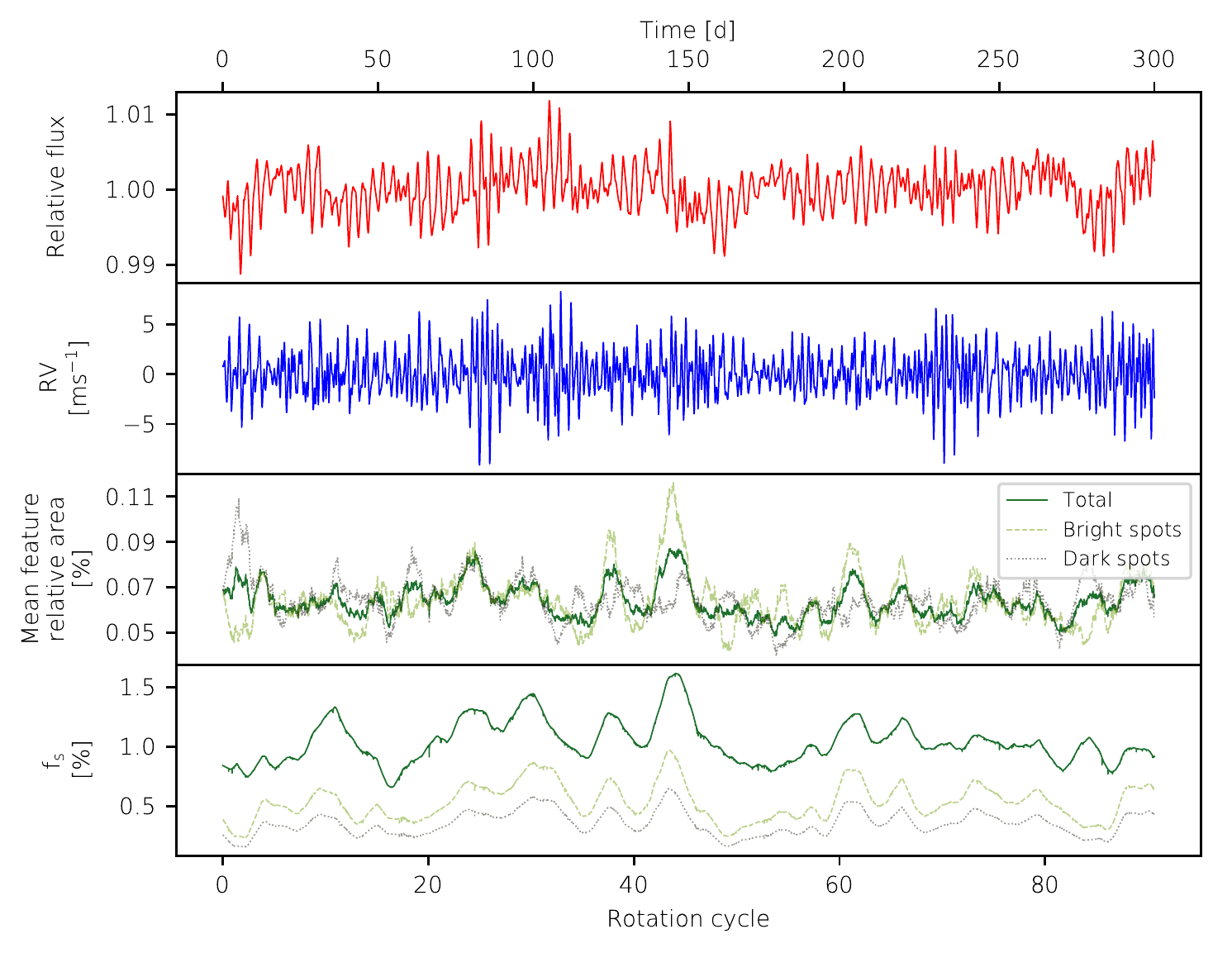}
    \caption{Synthetic RV and light curves of TRAPPIST-1: (\textit{top}) relative brightness curve with respect to the quiet photosphere; (\textit{second}) stellar activity RV curve}; (\textit{third}) Mean feature relative area with respect to the visible disk; (\textit{bottom}) f$_{\mathrm{s}}$ accounting for total surface covered by bright and/or dark features with respect to the whole stellar surface. In the two lower panels, the quantities relative to dark features are plotted in gray dotted lines, those relative to bright features are plotted is light-green dashed lines and the dark-green solid lines refer to both dark and bright features.
    \label{fig:simulated_act}
\end{figure*}

\begin{figure}
	\includegraphics[width=\columnwidth]{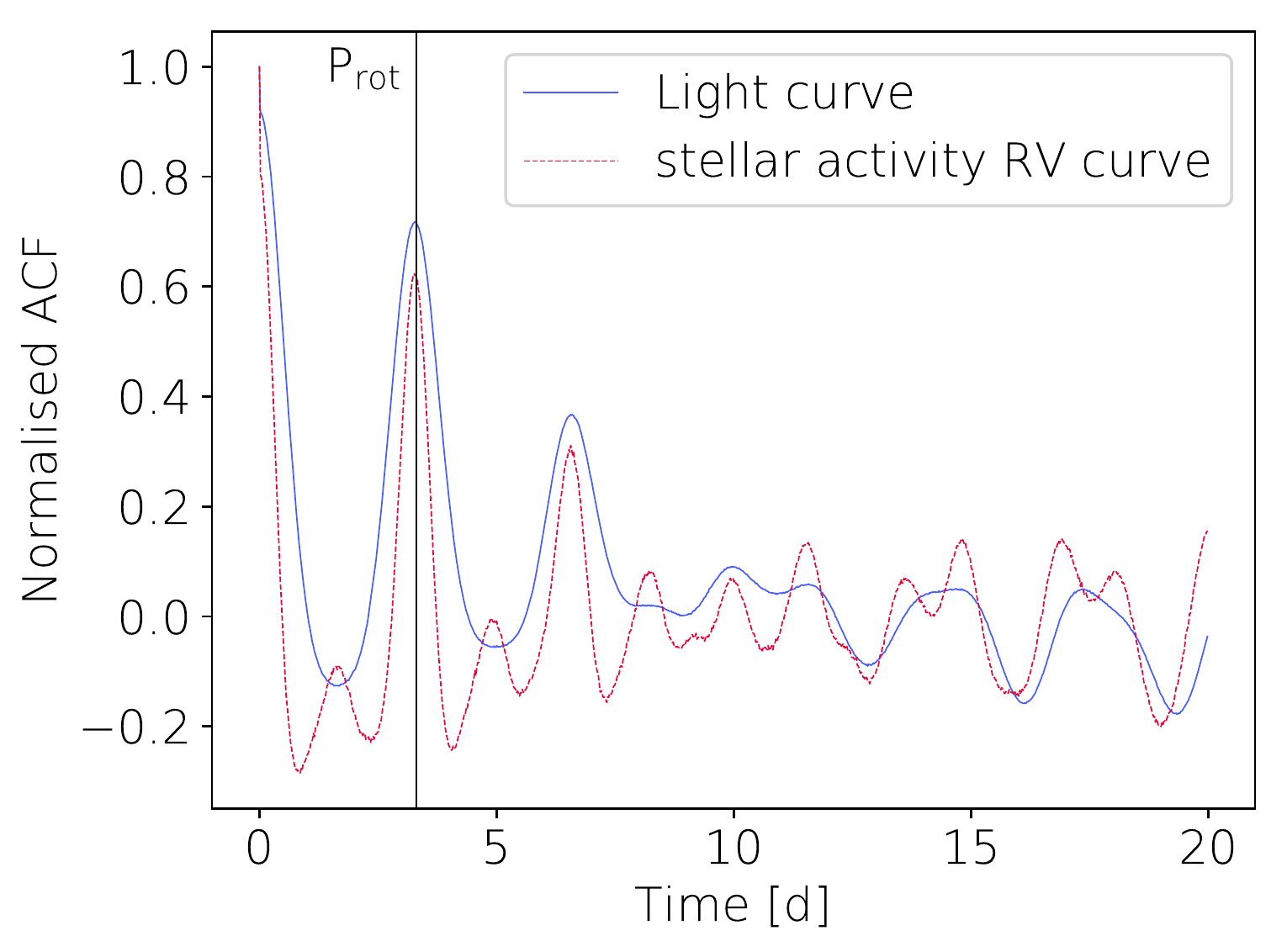}
    \caption{Normalized autocorrelation functions (ACF) of the noise-free stellar photometric (blue solid line) and RV curves (red dashed-line). Trappist-1 rotation period is indicated through the black vertical line.}
    \label{fig:acf}
\end{figure}

The synthetic photometric and RV activity curves are shown in Fig~\ref{fig:simulated_act}\footnote{The modeling of the stellar activity RV curve presented here ignores both the smaller brightness contrast of surface features at nIR wavelengths on the one hand, and the impact of the magnetic fields that these features likely host on the other hand;  preliminary simulations indicate that, whereas the first effect decreases the amplitude of the stellar activity RV curve, the second one amplifies it, with both effects more or less cancelling each other for magnetic field strengths of $\sim$2~kG.  We thus conclude that despite these simplifications, our simulation is able to predict the shape and amplitude of the stellar activity RV curve within better than a factor of 2.}. We obtain a stellar activity signal with an amplitude ranging from $\sim$5~\ms\ to almost 10~\ms\ depending on the epoch. The lower two panels respectively show the mean size and surface coverage of the bright and dark features in our model. The average feature relative area, $S_{\mathrm{r}}$~$\sim$~0.06\% of the total stellar area, is slightly larger than the features modeled in \citet{rackham2018} and lies close to threshold of undetected features in \citet{morris2018b}. The auto-correlation functions (ACFs) of the noise-less photometric and RV curves are shown in Fig~\ref{fig:acf}. Both time-series show 3.3~d-quasi-periodic shapes symptomatic of short-lived active regions with half-life of $\sim$2~P$_{\mathrm{rot}}$. The ACF peak at a difference of one P$_{\mathrm{rot}}$ is damped, though less than the one in the original K2 light curve \citep[e.g. Fig 6 in][]{morris2018a}, partly as a result of the noise not being included yet. 




\begin{figure}
	\includegraphics[width=\columnwidth]{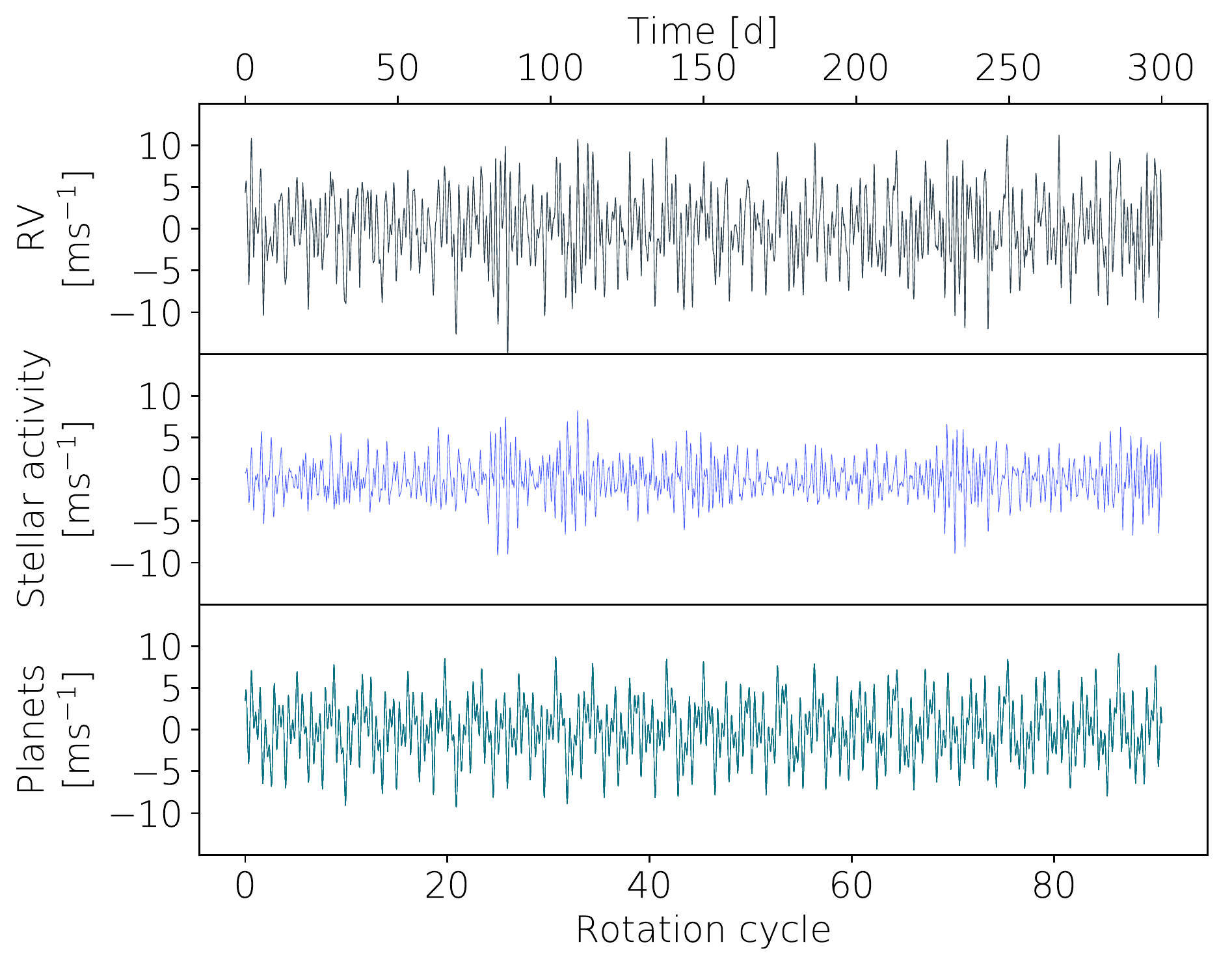}
    \caption{\textit{Top:} Synthetic TRAPPIST-1 RV curve containing both stellar activity signal and planetary signatures. \textit{Middle:} stellar activity RV signal shown in Fig \ref{fig:simulated_act}. \textit{Bottom:} Planetary signature containing the RV signal of the seven planets.}
    \label{fig:signal_tot}
\end{figure}


\subsection{Scheduling of the observations}

The synthetic TRAPPIST-1 RV curve (i.e. planet and stellar activity RV signals) is shown in Fig~\ref{fig:signal_tot}. The lower two panels respectively display the stellar activity and planet RV curves whose levels of fluctuation are similar. We simulated TRAPPIST-1 RV follow-ups using the various sampling schemes listed in Table \ref{tab:cases}. Each sampling strategy provides a set of epochs at which we select the data points from the densely-sampled RV curve containing both the planetary and stellar RV curves. All datasets are built from the same 90-d visibility window to ensure that all simulations feature the same realization of the stellar activity signal.

More specifically, sampling scheme A leads to an evenly sampled dataset with a sampling frequency of 2~d$^{-1}$, i.e. greater than the Nyquist frequency of the planet signal. Schemes B and C assume that TRAPPIST-1 observations are carried out from CFHT and TBL, when the star is visible at an airmass of $\la$1.5. To account for stochastic weather conditions, the observations are achieved with a probability of 0.85 and 0.5 at CFHT and TBL respectively. Scheme C restricts CFHT observations to bright time periods ($\sim$15~d centered on full Moon periods) whereas scheme B allows observations in bright and dark time periods, making scheme C more consistent with typical observational constraints at CFHT. Finally, scheme D is similar to scheme B except that we assume that the TBL is not available, leaving us with only one telescope for following-up TRAPPIST-1, this time with no constrain on the moon phase.

\renewcommand{\arraystretch}{1.2}
\begin{table}
\centering
\caption{\label{tab:cases}List of the observational sampling strategies considered in this study. Columns~(1) to~(3) show respectively the name of the configuration, the size of the dataset and the sampling frequency. Additional details on the selection scheme are added in column~(4).}
\begin{tabular}{cccc}
\hline
Scheme & N$_{\mathrm{pts}}$ & f$_{\mathrm{s}}$ & Comments \\
\hline
A & 180 & 2 d$^{-1}$ & Even sampling\\
B & 120 & $\sim$ 2 d$^{-1}$  & CFHT (bright \& dark times) \& TBL \\
C & 86 & $\sim$ 2 d$^{-1}$  &  CFHT (bright time only) \& TBL \\
D & 76 & $\sim$ 1 d$^{-1}$  & CFHT only (bright \& dark times) \\
\hline
\end{tabular}
\end{table}

\subsection{White noise}

We account for the instrumental and photon noises by adding random values drawn from a Gaussian distribution to each newly created dataset. For each sampling scheme listed in Table \ref{tab:cases}, we consider random noise levels of 1 and 2~\ms\ rms in order to quantify the losses in the precision of the mass estimates when the noise level increases. At 1~\ms, we assume the photon noise to be the dominant contribution of the white noise, implying thus that other sources of noise such as the instrument itself or the correction for tellurics in the spectra produce a contribution much smaller than 1~\ms\ which is optimistic and less realistic than the 2~\ms\ case. In what follows, each sampling scheme from Table \ref{tab:cases} will be indexed by the rms of the injected white noise expressed in \ms. For example, case A$_{1}$ designates a dataset sampled with scheme A with a white noise of 1~\ms\ rms.


\section{Fitting the RV time-series}\label{sec:section3}

We fit the synthetic RV time-series using the following model

\begin{eqnarray}
	\rm{V}_{\mathrm{r}}(t) = \rm{V}_{\mathrm{p}}(t) + \rm{V}_{\mathrm{j}}(t) + \epsilon (t)
\label{eq:sig}
\end{eqnarray}

\noindent 
where $\rm{V}_{\mathrm{p}}(t)$ and $\rm{V}_{\mathrm{j}}(t)$ respectively account for the planetary signature and stellar activity signal and $\epsilon(t)$~$\sim$~$\mathcal{N}(0,\sigma^{2}(t))$, where the error on the datapoint at time t, $\sigma(t)$, is assumed to be known. Furthermore, we assume that systematics, quantified with simultaneous observations of RV standards with SPIRou, are already corrected.

\subsection{Fit of the planetary signal only}
As a sanity check, we first consider the ideal case of a dataset containing only the planet signatures embedded in white noise, which can thus be modeled according to $\rm{V}_{\mathrm{r}}(t) = \rm{V}_{\mathrm{p}}(t) + \epsilon (t)$. We search for $\rm{N}_{\mathrm{p}}$ planets with circular orbits assuming no planet-planet interactions (see the justifications in Section~\ref{sec:section4}). Therefore, we fit the RV planet signatures using a linear combination of sinusoids such that

\begin{eqnarray}
\rm{V}_{\mathrm{p}}(t) = \sum_{n=1}^{N_{\mathrm{p}}} \alpha_{n} \cos \left( \frac{2 \pi}{\rm{P}_{n}} t \right) + \beta_{n} \sin \left( \frac{2 \pi}{\rm{P}_{n}} t \right)
\label{eq:pl_margin}
\end{eqnarray}

\noindent
where $\alpha_{\rm{n}}= \mathrm{K}_{\rm{n}} \cos \mathrm{\phi}_{\rm{n}}$ and $\beta_{n} = - \rm{K}_{n} \sin \mathrm{\phi}_{\rm{n}}$, with $\rm{K}_{n}$ and $\mathrm{\phi}_{\rm{n}}$ the semi-amplitude and orbital phase of the n-th planet. The planet orbital periods, P$_{\rm{n}}$, are frozen to their best estimates from \citetalias{delrez2018}. This results in a linear $2 \rm{N}_{\mathrm{p}}$-parameter model $\boldsymbol{\rm{V}_{\mathrm{p}}} = \boldsymbol{\rm{X}} \boldsymbol{\omega}$, where $\boldsymbol{\omega}~=~\left(\alpha_{1},\beta_{1},..,\alpha_{\rm{N}_{\mathrm{p}}}, \beta_{\rm{N}_{\mathrm{p}}} \right)$ and $\boldsymbol{\rm{X}}$ is a $\left(\rm{N}, 2\rm{N}_{\mathrm{p}} \right)$ array, $\rm{N}$ being the number of data points. Assuming a Gaussian white noise, the parameter posterior density function, $p(\boldsymbol{\omega} |\boldsymbol{\rm{V}_{\mathrm{r}}})$, can be analytically derived, leading to

\begin{eqnarray}
p(\boldsymbol{\omega} |\boldsymbol{\rm{V}_{\mathrm{r}}}) = \mathcal{\rm{N}} \left( \boldsymbol{\rm{A}}^{-1}\boldsymbol{\rm{b}}, \boldsymbol{\rm{A}}^{-1}  \right)
\label{eq:post_noact}
\end{eqnarray}

\noindent
where

\begin{eqnarray}
 \left\{
\begin{array}{lll}
  \boldsymbol{\rm{A}} & = & \boldsymbol{\rm{X}}^{\rm{T}} \boldsymbol{\Sigma}^{-1} \boldsymbol{\rm{X}} \\
  \boldsymbol{\rm{b}} & = & \boldsymbol{\rm{X}}^{\rm{T}} \boldsymbol{\Sigma}^{-1} \boldsymbol{\rm{V}_{\mathrm{r}}}
\end{array}
\right.
\label{eq:A_b_LS}
\end{eqnarray}

\noindent
where the covariance matrix of the white noise, $\Sigma$, is such that $\Sigma_{ij} = \sigma(t_{i})^{2} \delta_{ij}$, $\delta$ being the Kronecker delta.

We used Eq~\ref{eq:post_noact} to derive maximum a posteriori estimates as well as 1$\sigma$-uncertainties on the planet masses and orbital phases for the 4 sampling schemes listed in Table~\ref{tab:cases} at white noise levels of 1 and 2~\ms\ rms. In each case, the estimation is carried out on 50 signals with different white noise realizations from which we average the best planet parameters and error bars. 

In this ideal test case, we are able to recover the masses at a $>10\sigma$-precision level for planets b, c, g, f and e for all sampling schemes and white noise levels. The precision of the mass estimates for planets d and h systematically lies above 5 $\sigma$ except for cases C$_{2}$ and D$_{2}$ for which the mass of planet h is estimated at a precision $\ga$3~$\sigma$.

\subsection{Fitting the stellar activity RV signal}

We now fit a signal that contains no planet, but only the stellar activity signal and white noise. In this purpose, we use Gaussian Process Regression \citep[GPR][]{rasmussen2006} to model the stellar activity RV signal. The shape of the ACF of the synthetic dataset (see Fig~\ref{fig:acf}) suggests that a quasi-periodic covariance kernel will be adapted to reconstruct the synthetic stellar activity RV curve \citep[][hereafter \citetalias{haywood2014}]{haywood2014}:

\begin{eqnarray}
\mathrm{k}(t_{i},t_{j}) = \mathrm{\theta}_{1}^{2} \exp \left[ - \frac{(t_{i}-t_{j})^{2}}{\mathrm{\theta}_{2}^{2}} - \frac{\sin^{2} \frac{\pi (t_{i}-t_{j})}{\mathrm{\theta}_{3}}}{\mathrm{\theta}_{4}^{2}} \right]
\label{eq:cov_fct}
\end{eqnarray}

\noindent 
where $t_{i}$ and $t_{j}$ are the times associated with observations $i$ and $j$. This covariance function depends on four hyperparameters called $\theta_{1}$ to $\theta_{4}$; $\theta_{1}$ is the amplitude of the Gaussian Process (GP) which scales with the amplitude of the stellar activity signal, $\theta_{2}$ is the time-scale for the evolution of the active regions, $\theta_{3}$ is the period of the GP whereas, $\theta_{4}$, called the smoothing parameter, controls the number of high-frequency structures that can be included into the fit. In what follows, the vector containing the four hyperparameters is called $\boldsymbol{\theta}$.

We used GPR to independently model photometric and RV time-series of the stellar activity signal evenly sampled with scheme A. White noises of 1 \ms\ and 1 mmag rms were added respectively to the RV and photometric time-series so that the two datasets have similar signal-to-noise ratios. The 4 GP hyperparameters are jointly estimated by maximising the posterior density function of the hyperparameters, $p(\boldsymbol{\theta}|\boldsymbol{\rm{V}_{\mathrm{r}}})$, sampled using a Bayesian Markov Chain Monte Carlo (MCMC). We adopted the non-informative priors listed in Table \ref{tab:priors} to ensure a physically realistic fit of the stellar activity signal. The period of the GP is expected to lie close to the stellar rotation period \citep{angus2018}, explaining thus the narrow prior distribution adopted for $\theta_{3}$. The likelihood of the GP fit to the stellar activity RV signal, $\mathcal{L} = p(\boldsymbol{\rm{V}_{\mathrm{r}}}|\boldsymbol{\theta})$, is such that

\begin{eqnarray}
2 \ln \mathcal{L} = -\rm{N} \ln{2 \pi} - \ln \left( |\boldsymbol{\rm{K}}+\boldsymbol{\Sigma}| \right) - \boldsymbol{\rm{V}_{\mathrm{r}}}^{\rm{T}} \left( \boldsymbol{\rm{K}}+\boldsymbol{\Sigma} \right)^{-1} \boldsymbol{\rm{V}_{\mathrm{r}}}
\label{eq:likelihood}
\end{eqnarray}

\noindent
where $\boldsymbol{\rm{K}}$ is the GP covariance matrix. Concretely, we use the \textsc{emcee} affine invariant sampler \citep[][100 walkers, 2000 iterations]{Foreman-Mackey2013} to sample the posterior density. After removing a burn-in period of $\sim$10 autocorrelation times from the chain, we compute the median and uncertainties on each parameter from the posterior distribution\footnote{The 1$\sigma$ error bars are given by $0.5(x_{2}-x_{1})$, where $x_{1}$ and $x_{2}$ respectively correspond to the 16$^{\rm{th}}$ and 84$^{\rm{th}}$ percentiles of the posterior distributions of each parameter.}


\begin{table}
\centering
\caption{\label{tab:priors}Prior density probabilities used for the MCMC sampling for the interpolation of the stellar activity.}
\begin{tabular}{cc}
\hline
Parameter & Prior \\
\hline
$\ln \theta_{1}$ [$\ln$(\ms)] & $\mathcal{U}(-1,10)$ \\
$\ln \theta_{2}$ [$\ln$(d)] & $\mathcal{U}(0,5)$ \\
$\theta_{3}$ [P$_{\text{rot}}$] &  $\mathcal{U}(0.9,1.1)$ \\
$\theta_{4}$ & $\mathcal{U}(0.1,4.0)$ \\
\hline
\end{tabular}
\end{table}

\begin{figure}
\centering
\includegraphics[width=\columnwidth,height=9cm]{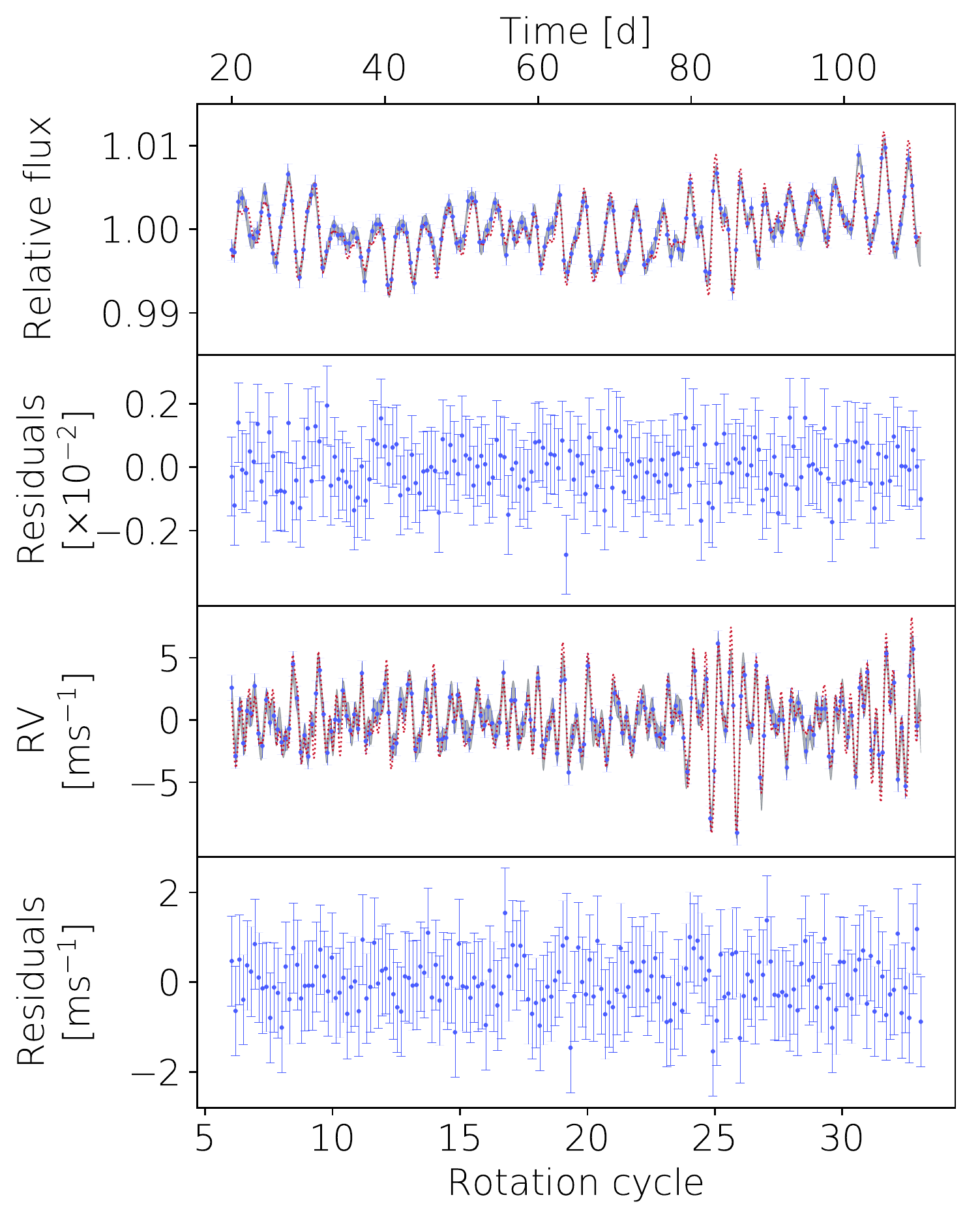}
\caption{From top to bottom: Photometric time series, residuals after the GP interpolation of the photometric time-series, RV time series and residuals after the GPR. The red dotted-lines are the injected signals and the blue dots with error bars are the GP prediction at the observation times. The gray error bands are the $\pm$1~$\sigma$ predictions of the GP.}
\label{fig:fit_actonly}
\end{figure}

This process is repeated for 50 photometric and RV time-series with different white noise realizations, resulting in residuals of about 0.6~\ms\ and 0.8~mmag rms for RV and photometric time-series respectively. Table~\ref{tab:results_actonly} shows the average over the 50 time-series of the best and the median values of the hyperparameters. We find similar GP periods and timescales for both photometric and RV time-series, consistent with the injected rotation period and the one derived from the ACF of Fig~\ref{fig:acf}. We note that $\theta_{4}$ is roughly twice as large in photometry as in RV, reflecting the fact that a given feature at the stellar surface produces a signature in the RV curve that evolves twice as fast as its counterpart in the lightcurve. The interpolation of photometric and RV time-series is shown in Fig~\ref{fig:fit_actonly} for a given realization of white noise.

\begin{table}
\centering
\renewcommand{\arraystretch}{1.3}
\caption{\label{tab:results_actonly}Best and median estimates of the four hyperparameters of the GP when fitting the light and stellar activity RV curves without any planetary component. Note that the values indicated in this table are averaged over 50 datasets with different white noise realizations of 1~\ms\ and 1~mmag for the RV and photometric curves respectively.}
\begin{tabular}{c|cc|cc}
\hline
Parameter & \multicolumn{2}{c}{Lightcurve} & \multicolumn{2}{c}{RV} \\
& Best & Median & Best & Median \\ \hline
$\theta_{1}$ & 0.33 \% & $0.34 \pm 0.05$ \% & 2.6 \ms\ & $2.6 \pm 0.3$ \ms \\
$\theta_{2}$ [d] & 7.2 & $7.2 \pm 0.8$ & 7.9 & $7.8 \pm 0.8$ \\
$\theta_{3}$ [d] & 3.30 & $3.30 \pm 0.04$ & 3.30 & $3.30 \pm 0.02$ \\
$\theta_{4}$ & 0.65 & $0.67 \pm 0.09$ & 0.37 & $0.37 \pm 0.03$ \\
\hline
\end{tabular}
\renewcommand{\arraystretch}{1.2}
\end{table}

Following the approach of \citet{aigrain2012}, we computed the first derivative of the synthetic noise-free lightcurve and found a linear correlation coefficient of -0.89 with the stellar activity RV curve, indicating a net anti-correlation between the two signals. We fitted the evenly sampled stellar activity RV curve from Fig~\ref{fig:fit_actonly} with the first derivative of the stellar lightcurve, evaluated at the same time values, using a linear model. This resulted in residuals of 1.4~\ms\ rms, i.e. larger than that obtained when the stellar activity signal is modeled with GP (0.56~\ms\ rms).


\subsection{Simultaneous fit of stellar activity RV signal and planetary curves}\label{sec:3.3}

We now model the synthetic datasets described in Section~\ref{sec:section2} using Eq~\ref{eq:sig}, including all model components (planet, stellar activity and white noise). The goal is to simultaneously filter the stellar activity signal with a GP directly trained in the dataset and estimate the planetary semi-amplitudes and orbital phases. This time, our model contains $2 \rm{N}_{\mathrm{p}}$ linear planet parameters and 4 non-linear GP hyperparameters. We use a Bayesian MCMC to sample the posterior density marginalised over the planet parameters:

\begin{eqnarray}
 p(\boldsymbol{\theta} | \boldsymbol{V_{\mathrm{r}}} ) \propto  \pi(\boldsymbol{\theta}) \int_{\boldsymbol{\omega}} p(\boldsymbol{V_{\mathrm{r}}} | \boldsymbol{\theta}, \boldsymbol{\omega})  \pi(\boldsymbol{\omega}) d \boldsymbol{\omega}
\label{eq:post_act_pl}
\end{eqnarray}


\noindent
where $\boldsymbol{\theta}$ and $\boldsymbol{\omega}$ are assumed independent, and where $\pi(\boldsymbol{\theta})$ and $\pi(\boldsymbol{\omega})$ stand for the prior density on $\boldsymbol{\theta}$ (listed in Table~\ref{tab:priors}) and the linearised planetary amplitudes respectively. We analytically derive Eq~\ref{eq:post_act_pl} assuming an infinitely broad Gaussian prior density for $\boldsymbol{\omega}$ which is similar as a non-informative uniform prior density \citep[see Section 2.7.1 in][for a detailed generic demonstration and expression for the likelihood]{rasmussen2006}. After removing a burn-in period of a few autocorrelation times from the chain, we compute the set of GP hyperparameters, $\boldsymbol{\theta_{\rm{best}}}$, that maximizes the posterior density as well as the median and error bars on the hyperparameters from the posterior samples. At a given posterior sample $\boldsymbol{\theta}$, the posterior distribution for the planet parameters is such that:





\begin{eqnarray}
p(\boldsymbol{\omega} |\boldsymbol{\rm{V}_{\mathrm{r}}}, \boldsymbol{\theta}) =  \mathcal{N} \left( \boldsymbol{\rm{A}}(\boldsymbol{\theta})^{-1}\boldsymbol{\rm{b}}(\boldsymbol{\theta}), \boldsymbol{\rm{A}}(\boldsymbol{\theta})^{-1}  \right)
\label{eq:post_pl_act_fin}
\end{eqnarray}

\noindent
where

\begin{eqnarray}
 \left\{
\begin{array}{lll}
  \boldsymbol{\rm{A}}(\boldsymbol{\theta}) & = & \boldsymbol{\rm{X}}^{\rm{T}} \left[ \boldsymbol{\Sigma} + \boldsymbol{\rm{K}}(\boldsymbol{\theta}) \right]^{-1} \boldsymbol{X} \\
  \boldsymbol{\rm{b}}(\boldsymbol{\theta}) & = & \boldsymbol{\rm{X}}^{\rm{T}} \left[ \boldsymbol{\Sigma} + \boldsymbol{\rm{K}}(\boldsymbol{\theta}) \right]^{-1} \boldsymbol{\rm{V}_{\mathrm{r}}}
\end{array}
\right.
\label{eq:A_b_LS_act}
\end{eqnarray}



\noindent
We use Eq~\ref{eq:post_pl_act_fin} at $\boldsymbol{\theta} = \boldsymbol{\theta_{\rm{best}}}$ to estimate the planet parameters as well as their $1\sigma$-error bars. Note that these error bars are significantly larger than the dispersion of the best planet parameters obtained by computing Eq~\ref{eq:post_pl_act_fin} for all the samples of the joint posterior distribution. The planetary masses and their error bars are finally derived from the estimates of the semi-amplitudes using Kepler's third law. We stress that this procedure is in principle applicable to any planetary system with circular orbits of known periods (and phases).


\subsection{Model comparison and evidence estimation}

Which of the TRAPPIST-1 planets are detected with the algorithm described above? Although apparently intuitive, this issue requires the use of a robust criterion to be addressed in a reliable way. In our case, we carried out a Bayesian comparison of models searching respectively for $N_{\mathrm{p}}$ and $N_{\mathrm{p}}+1$ planets, where $N_{\mathrm{p}}$ varies from 0 to 6. Assuming that we have no prior information on the number of searched planets, two models are compared by computing the so-called Bayes factor, i.e. the ratio of the Marginal Likelihoods (MLs) of the two models. The ML being mathematically intractable in our case, we used the method of \citet[][hereafter \citetalias{Chib2001}]{Chib2001} to approximate its value from the posterior density obtained when sampling the parameter space of each model \citepalias[see also][for more details on the implementation of the method]{haywood2014}.

For each dataset, we compute the ML of models searching for 0 to 7 planets, where the searched planets are added by decreasing order of RV semi-amplitude (i.e. TRAPPIST-1 b, c, g, f, e, d and h). We then derive the Bayes factor comparing every model containing at least one planet with the model searching for one fewer planet. According to \citet{Jeffreys1961}, a Bayes factor greater than 150 ($\sim$5 in log) is associated to a fair detection of the planet added to the model. The evidence in favor of the planet will be regarded as strong if the Bayes factor lies within 150-20 ($\sim$5-3 in log), positive if it falls within the range 20-3 ($\sim$3-1 in log) and inconclusive otherwise.

\section{Results}\label{sec:section4}

\subsection{Fitting simulated data}

We used the model described in Section~\ref{sec:section3} to recover TRAPPIST-1 planet masses for the sampling schemes listed in Table \ref{tab:cases} and for white noise levels of 1 and 2~\ms\ rms. Our first goal being to assess the effects of the stellar activity signal and sampling strategies on the quality of the mass estimates, we start by systematically searching for 7 planets in the datasets. For each sampling scheme and white noise level, we fit 50 time-series, each with different white noise realisations of equal rms. The best estimates of the planet masses and GP hyperparameters with error bars are obtained by taking the median of the distributions. The outcome of the MCMC samplings and the best estimates for the planet masses are shown in Tables \ref{tab:results_tot} and \ref{tab:masses} respectively.

\begin{table*}
\centering
\caption{Results of the interpolation of the synthetic datasets for the sampling schemes listed in Table \ref{tab:cases} and for white noise levels of 1 and 2~\ms. Column (1) indicates the sampling case indexed by the rms of the injected white noise expressed in \ms, column (2) is the number of searched planets and columns (3) to (6) are the medians and 1~$\sigma$ error bars of the GP hyperparameters averaged over $\sim$50 input datasets with different realisations of white noise. We show the logarithm of the maximum likelihood, marginal likelihood and Bayes factor for each model compared to the model with one less planet in columns (7), (8) and (9). The median reduced $\chi^{2}$ of the fit, $\chi_{\rm{r}}^{2}$ (i.e.~$\chi^{2}/\rm{N}$ where $\rm{N}$ is the number of data points), is shown in column (10).} The hyperparameters listed without error bars were frozen to the indicated value during the MCMC sampling to ensure the convergence of the algorithm.\label{tab:results_tot}
\begin{tabular}{c|c|cccc|cccc}
Case & N$_{\mathrm{p}}$ & $\theta_{1}$ & $\theta_{2}$ & $\theta_{3}$ & $\theta_{4}$  & ln($\mathcal{L}_{\rm{max}}$) & ln(ML) & Bayes factor & $\chi_{\rm{r}}^{2}$ \\
 & & [ms$^{-1}$] & [d] & [d] & & & & [$\rm{N}_{p}$;$\rm{N}_{p}-1$] & \\
\hline
\multirow{3}{*}{A$_{1}$} & 3 & 2.94 $\pm$ 0.22 & 23.51 $\pm$ 12.20 & 3.25 $\pm$ 0.04 & 0.10 $\pm$ 0.07 & -439.7 $\pm$ 3.2 & -439.1 $\pm$ 3.9 & -- & 0.1\\
 & 4 & 2.77 $\pm$ 0.23 & 7.98 $\pm$ 0.74 & 3.27 $\pm$ 0.01 & 0.24 $\pm$ 0.02 & -421.3 $\pm$ 3.5 & -421.8 $\pm$ 3.4 & 17.6 $\pm$ 3.2 & 0.2\\
 & 5 & 2.69 $\pm$ 0.27 & 7.53 $\pm$ 0.76 & 3.30 $\pm$ 0.02 & 0.38 $\pm$ 0.04 & -393.9 $\pm$ 4.1 & -394.6 $\pm$ 4.1 & 27.0 $\pm$ 2.1 & 0.3\\
 & 6 & 2.64 $\pm$ 0.26 & 7.70 $\pm$ 0.80 & 3.30 $\pm$ 0.02 & 0.37 $\pm$ 0.04 & -391.8 $\pm$ 4.1 & -392.6 $\pm$ 4.1 & 2.2 $\pm$ 0.8 & 0.3\\
 & 7 & 2.64 $\pm$ 0.27 & 7.70 $\pm$ 0.81 & 3.30 $\pm$ 0.02 & 0.37 $\pm$ 0.04 & -391.5 $\pm$ 4.1 & -392.2 $\pm$ 4.1 & 0.5 $\pm$ 0.4 & 0.3\\
\hline
\multirow{3}{*}{A$_{2}$} & 3 & 2.89 $\pm$ 0.27 & 20.16 $\pm$ 12.60 & 3.26 $\pm$ 0.09 & 0.12 $\pm$ 0.09 & -469.1 $\pm$ 5.2 & -469.6 $\pm$ 5.6 & -- & 0.4\\
 & 4 & 2.73 $\pm$ 0.28 & 9.04 $\pm$ 2.79 & 3.28 $\pm$ 0.05 & 0.23 $\pm$ 0.06 & -459.0 $\pm$ 5.5 & -459.8 $\pm$ 5.7 & 10.3 $\pm$ 3.0 & 0.4\\
 & 5 & 2.61 $\pm$ 0.30 & 7.64 $\pm$ 1.74 & 3.29 $\pm$ 0.05 & 0.34 $\pm$ 0.06 & -445.9 $\pm$ 4.9 & -446.8 $\pm$ 4.9 & 13.0 $\pm$ 2.8 & 0.5\\
 & 6 & 2.58 $\pm$ 0.31 & 7.78 $\pm$ 1.77 & 3.28 $\pm$ 0.04 & 0.34 $\pm$ 0.06 & -444.0 $\pm$ 4.8 & -444.9 $\pm$ 4.8 & 2.2 $\pm$ 1.1 & 0.5\\
 & 7 & 2.59 $\pm$ 0.31 & 7.82 $\pm$ 1.80 & 3.28 $\pm$ 0.04 & 0.34 $\pm$ 0.06 & -443.3 $\pm$ 4.6 & -444.2 $\pm$ 4.6 & 1.0 $\pm$ 0.7 & 0.5\\
\hline
\multirow{3}{*}{B$_{1}$} & 3 & 3.09 $\pm$ 0.27 & 9.13 $\pm$ 3.95 & 3.28 $\pm$ 0.06 & 0.16 $\pm$ 0.09 & -298.7 $\pm$ 2.9 & -299.1 $\pm$ 2.9 & -- & 0.1\\
 & 4 & 2.82 $\pm$ 0.26 & 8.50 $\pm$ 1.84 & 3.26 $\pm$ 0.03 & 0.20 $\pm$ 0.04 & -286.6 $\pm$ 3.0 & -287.4 $\pm$ 3.0 & 11.8 $\pm$ 1.5 & 0.2\\
 & 5 & 2.76 $\pm$ 0.32 & 8.42 $\pm$ 1.17 & 3.29 $\pm$ 0.03 & 0.37 $\pm$ 0.05 & -268.9 $\pm$ 3.6 & -269.7 $\pm$ 3.6 & 18.0 $\pm$ 2.2 & 0.2\\
 & 6 & 2.76 $\pm$ 0.32 & 8.66 $\pm$ 1.24 & 3.28 $\pm$ 0.03 & 0.37 $\pm$ 0.05 & -267.6 $\pm$ 3.4 & -268.5 $\pm$ 3.4 & 1.4 $\pm$ 0.8 & 0.2\\
 & 7 & 2.73 $\pm$ 0.32 & 8.96 $\pm$ 1.35 & 3.28 $\pm$ 0.03 & 0.37 $\pm$ 0.05 & -265.7 $\pm$ 3.5 & -266.5 $\pm$ 3.5 & 2.0 $\pm$ 0.9 & 0.2\\
\hline
\multirow{3}{*}{B$_{2}$} & 3 & 3.08 $\pm$ 0.34 & 11.15 $\pm$ 6.42 & 3.34 $\pm$ 0.11 & 0.16 $\pm$ 0.13 & -316.2 $\pm$ 3.5 & -317.0 $\pm$ 3.4 & -- & 0.3\\
 & 4 & 2.83 $\pm$ 0.34 & 8.59 $\pm$ 3.36 & 3.26 $\pm$ 0.08 & 0.23 $\pm$ 0.09 & -307.2 $\pm$ 3.3 & -308.2 $\pm$ 3.3 & 8.9 $\pm$ 2.0 & 0.4\\
 & 5 & 2.74 $\pm$ 0.37 & 8.21 $\pm$ 2.22 & 3.27 $\pm$ 0.06 & 0.35 $\pm$ 0.08 & -298.3 $\pm$ 3.1 & -299.3 $\pm$ 3.2 & 9.1 $\pm$ 2.5 & 0.4\\
 & 6 & 2.74 $\pm$ 0.38 & 8.53 $\pm$ 2.35 & 3.27 $\pm$ 0.05 & 0.35 $\pm$ 0.08 & -297.0 $\pm$ 3.0 & -297.9 $\pm$ 3.0 & 1.6 $\pm$ 0.9 & 0.4\\
 & 7 & 2.72 $\pm$ 0.38 & 8.81 $\pm$ 2.53 & 3.27 $\pm$ 0.05 & 0.35 $\pm$ 0.08 & -295.3 $\pm$ 3.0 & -296.2 $\pm$ 3.0 & 1.9 $\pm$ 0.9 & 0.4\\
\hline
\multirow{3}{*}{C$_{1}$} & 3 & 3.32 $\pm$ 0.33 & 9.39 $\pm$ 4.42 & 3.29 $\pm$ 0.04 & 0.16 $\pm$ 0.09 & -220.3 $\pm$ 2.1 & -220.8 $\pm$ 2.1 & -- & 0.1\\
 & 4 & 3.16 $\pm$ 0.34 & 9.01 $\pm$ 3.17 & 3.28 $\pm$ 0.03 & 0.22 $\pm$ 0.06 & -213.1 $\pm$ 2.4 & -213.9 $\pm$ 2.4 & 6.9 $\pm$ 1.0 & 0.1\\
 & 5 & 2.96 $\pm$ 0.36 & 8.10 $\pm$ 1.91 & 3.31 $\pm$ 0.04 & 0.36 $\pm$ 0.06 & -201.6 $\pm$ 2.2 & -202.3 $\pm$ 2.2 & 11.8 $\pm$ 1.5 & 0.1\\
 & 6 & 2.90 $\pm$ 0.36 & 8.86 $\pm$ 2.22 & 3.30 $\pm$ 0.04 & 0.36 $\pm$ 0.06 & -198.6 $\pm$ 2.3 & -199.4 $\pm$ 2.2 & 3.0 $\pm$ 0.8 & 0.2\\
 & 7 & 2.90 $\pm$ 0.36 & 8.74 $\pm$ 2.20 & 3.29 $\pm$ 0.04 & 0.36 $\pm$ 0.06 & -197.8 $\pm$ 2.1 & -198.5 $\pm$ 2.1 & 0.9 $\pm$ 0.4 & 0.2\\
\hline
\multirow{3}{*}{C$_{2}$} & 3 & 3.35 $\pm$ 0.42 & 11.94 $\pm$ 7.15 & 3.29 $\pm$ 0.11 & 0.17 $\pm$ 0.10 & -230.2 $\pm$ 2.5 & -231.1 $\pm$ 2.4 & -- & 0.3\\
 & 4 & 3.18 $\pm$ 0.43 & 10.35 $\pm$ 5.48 & 3.28 $\pm$ 0.09 & 0.24 $\pm$ 0.09 & -224.4 $\pm$ 2.0 & -225.3 $\pm$ 2.0 & 5.9 $\pm$ 1.2 & 0.3\\
 & 5 & 2.91 $\pm$ 0.43 & 8.79 $\pm$ 3.90 & 3.29 $\pm$ 0.10 & 0.34 $\pm$ 0.10 & -216.4 $\pm$ 2.8 & -217.3 $\pm$ 2.9 & 8.0 $\pm$ 1.8 & 0.4\\
 & 6 & 2.86 $\pm$ 0.44 & 9.80 $\pm$ 4.59 & 3.26 $\pm$ 0.10 & 0.34 $\pm$ 0.10 & -213.6 $\pm$ 2.5 & -214.5 $\pm$ 2.5 & 3.0 $\pm$ 1.4 & 0.4\\
 & 7 & 2.88 $\pm$ 0.46 & 10.91 $\pm$ 5.51 & 3.24 $\pm$ 0.10 & 0.33 $\pm$ 0.11 & -212.3 $\pm$ 2.7 & -213.2 $\pm$ 2.8 & 1.4 $\pm$ 0.6 & 0.4\\
\hline
\multirow{3}{*}{D$_{1}$} & 3 & 3.25 $\pm$ 0.34 & 7.9 & 3.43 $\pm$ 0.04 & 0.37 & -190.1 $\pm$ 2.3 & -191.2 $\pm$ 2.3 & -- & 0.1\\
 & 4 & 3.01 $\pm$ 0.33 & 7.9 & 3.42 $\pm$ 0.04 & 0.37 & -184.2 $\pm$ 2.3 & -185.3 $\pm$ 2.3 & 5.9 $\pm$ 0.8 & 0.1\\
 & 5 & 2.59 $\pm$ 0.31 & 7.9 & 3.31 $\pm$ 0.05 & 0.37 & -173.3 $\pm$ 2.5 & -174.4 $\pm$ 2.5 & 10.9 $\pm$ 1.4 & 0.2\\
 & 6 & 2.60 $\pm$ 0.32 & 7.9 & 3.31 $\pm$ 0.05 & 0.37 & -172.3 $\pm$ 2.5 & -173.4 $\pm$ 2.5 & 1.2 $\pm$ 0.5 & 0.2\\
 & 7 & 2.47 $\pm$ 0.32 & 7.9 & 3.30 $\pm$ 0.05 & 0.37 & -168.7 $\pm$ 2.7 & -169.8 $\pm$ 2.7 & 3.7 $\pm$ 0.7 & 0.2\\
\hline
\multirow{3}{*}{D$_{2}$} & 3 & 3.03 $\pm$ 0.48 & 7.9 & 3.43 $\pm$ 0.07 & 0.37 & -199.0 $\pm$ 3.4 & -200.1 $\pm$ 3.4 & -- & 0.4\\
 & 4 & 2.81 $\pm$ 0.50 & 7.9 & 3.40 $\pm$ 0.09 & 0.37 & -194.1 $\pm$ 3.7 & -195.3 $\pm$ 3.7 & 5.1 $\pm$ 1.2 & 0.4\\
 & 5 & 2.52 $\pm$ 0.49 & 7.9 & 3.30 $\pm$ 0.11 & 0.37 & -187.6 $\pm$ 3.3 & -188.7 $\pm$ 3.3 & 6.9 $\pm$ 2.1 & 0.4\\
 & 6 & 2.56 $\pm$ 0.50 & 7.9 & 3.29 $\pm$ 0.11 & 0.37 & -186.5 $\pm$ 3.2 & -187.6 $\pm$ 3.1 & 1.1 $\pm$ 0.5 & 0.4\\
 & 7 & 2.41 $\pm$ 0.51 & 7.9 & 3.29 $\pm$ 0.12 & 0.37 & -182.9 $\pm$ 3.3 & -184.1 $\pm$ 3.3 & 3.6 $\pm$ 1.0 & 0.4\\
\hline
\end{tabular}
\end{table*}

\begin{figure*}
\includegraphics[width=\linewidth]{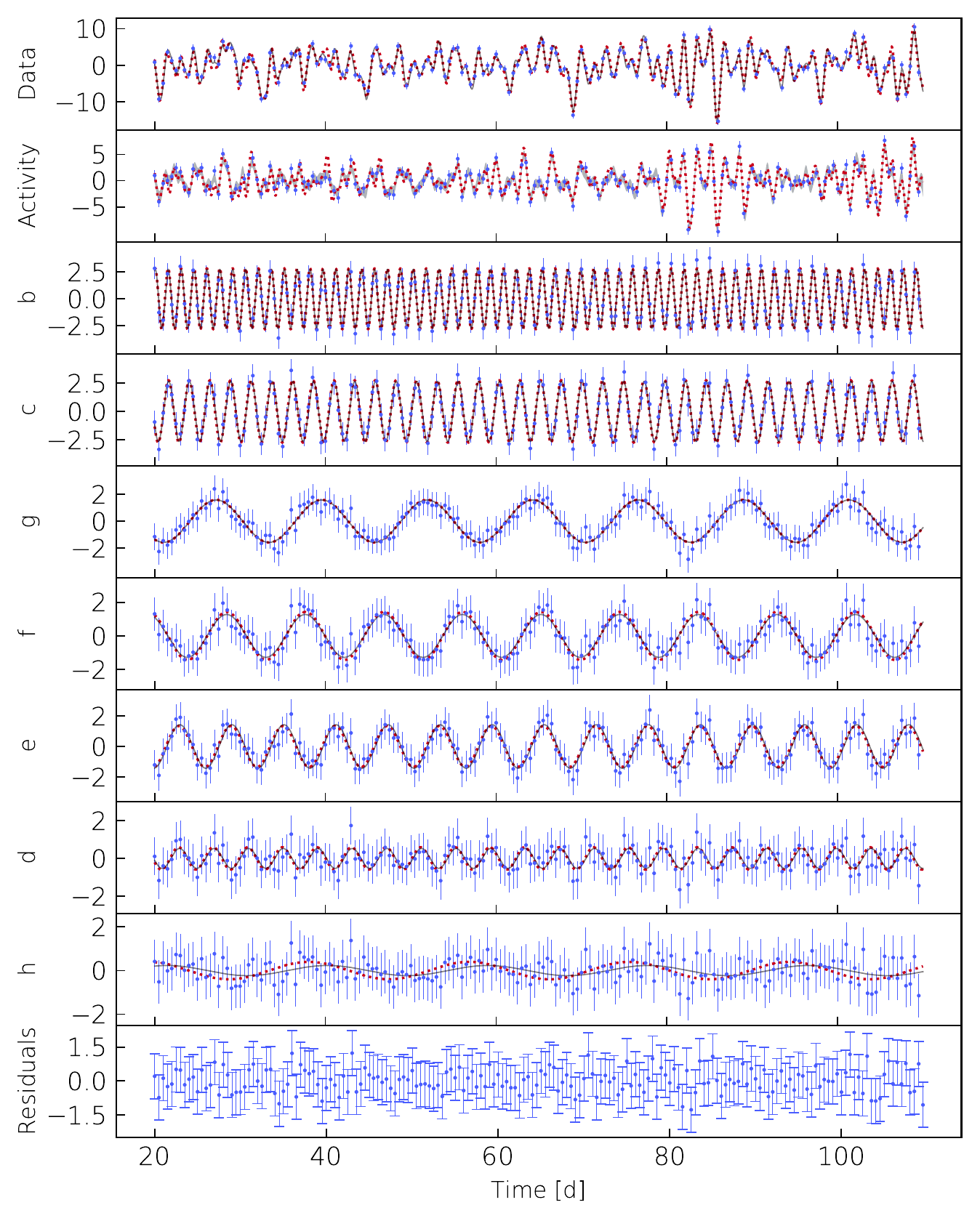}
\caption{Fit of a dataset sampled with A$_{1}$ scheme with a given realization of the 1~\ms\ white noise. All RVs are expressed in \ms. From top to bottom: raw dataset, stellar activity RV signal, RV signatures from planets b to h in a decreasing order of semi-amplitude and residuals (rms of 0.5~\ms). In each panel, the red dotted line is the injected RV curve and the grey solid line is the best prediction from the model. For the stellar activity and planetary signals, the blue dots are obtained by subtracting all the components of the model except the one displayed in the data.}
\label{fig:prediction_A1}
\end{figure*}

Case A$_{1}$, i.e. evenly sampled data with 1~\ms\ white noise, provides the best fit of the signal with an average reduced $\chi^{2}$, $\chi_{\rm{r}}^{2}$ (i.e. $\chi^{2}/\rm{N}$, where N is the number of data points), of $\sim$0.3. We find hyperparameters consistent with those obtained when fitting the stellar activity signal alone (see Table \ref{tab:results_actonly}), indicating a good interpolation of the stellar activity signal by the GP. Compared to the mass estimates obtained with the TTVs, our mass estimates are significantly more precise for planets b and c, and show comparable precision for planets g, f and e. However, planets d and h whose masses were derived with $>$10~$\sigma$ precision in the absence of stellar activity in case A$_{1}$, now show largely degraded mass estimates, with a precision level below 3~$\sigma$. The error bars on the planet orbital phase estimates are $\sim$3 to $\sim$140 times larger than the phase-shift induced by the maximum TTVs predicted from \citetalias{grimm2018} models (see Table~\ref{tab:planets}), justifying thus to have ignored the dynamical interactions between the planets when modelling the data. An example of the fit to our A$_{1}$ dataset is shown in Fig~\ref{fig:prediction_A1}, where we displayed, in particular, the reconstructed activity and planetary RV curves as well as the residuals of the fit to the synthetic data. The corresponding posterior distribution of the four GP hyperparameters is shown in Fig \ref{fig:post_A1}. In Appendix \ref{app:A}, we show examples of reconstructed RV times series as well as posterior densities for all sampling schemes with 1 and 2~\ms\ rms white noise.

When we increase the white noise level from 1 to 2~\ms\ (i.e. case~\Aa), we note that the error bars on the hyperparameter estimates are roughly doubled except for the amplitude of the GP whose precision is only slightly degraded. Compared to case A$_{1}$, the error bars on the mass estimates of planets c, f and e are multiplied by a factor $\sim$2, implying that the main limitation comes from the white noise itself. In contrast, the precision on the mass estimates of planets b, d and h is dominated by the stellar activity signal itself, as evidenced by the low increase in the error bars when switching from case A$_{1}$ to A$_{2}$. With a white noise of 2~\ms, only the mass estimates of planets b and c remain more precise that those derived from TTVs.


\renewcommand{\arraystretch}{1.4}
\begin{table*}
\centering
\caption{Best TRAPPIST-1 planetary system mass estimates for datasets sampled with the schemes listed in Table \ref{tab:cases} at a white noise of 1 or 2 \ms\ rms. In each case, the values are averaged over $\sim$50 RV time-series with different white noise realisation. Case TTV stands for the best mass estimates from \citetalias{grimm2018}.\label{tab:masses}}
\begin{tabular}{c|ccccccc}
Case & m$_{\mathrm{b}}$ & m$_{\mathrm{c}}$ & m$_{\mathrm{g}}$ & m$_{\mathrm{f}}$ & m$_{\mathrm{e}}$ & m$_{\mathrm{d}}$ & m$_{\mathrm{h}}$ \\
 & [M$_{\earth}$] & [M$_{\earth}$] & [M$_{\earth}$] & [M$_{\earth}$] & [M$_{\earth}$] & [M$_{\earth}$] & [M$_{\earth}$] \\
\hline
TTV & 1.017$^{+0.154}_{-0.143}$ & 1.156$^{+0.142}_{-0.131}$ & 1.148$^{+0.098}_{-0.049}$ & 0.934$^{+0.08}_{-0.078}$ & 0.772$^{+0.079}_{-0.075}$ & 0.297$^{+0.039}_{-0.035}$ & 0.331$^{+0.056}_{-0.049}$ \\
\hline
A$_{1}$ & 0.988 $\pm$ 0.083 & 1.173 $\pm$ 0.049 & 1.130 $\pm$ 0.113 & 0.943 $\pm$ 0.077 & 0.762 $\pm$ 0.071 & 0.325 $\pm$ 0.133 & 0.332 $\pm$ 0.260 \\
A$_{2}$ & 0.991 $\pm$ 0.109 & 1.194 $\pm$ 0.093 & 1.116 $\pm$ 0.175 & 0.907 $\pm$ 0.147 & 0.773 $\pm$ 0.131 & 0.331 $\pm$ 0.155 & 0.402 $\pm$ 0.292 \\
B$_{1}$ & 1.040 $\pm$ 0.096 & 1.221 $\pm$ 0.079 & 1.127 $\pm$ 0.147 & 0.965 $\pm$ 0.124 & 0.837 $\pm$ 0.102 & 0.251 $\pm$ 0.129 & 0.565 $\pm$ 0.261 \\
B$_{2}$ & 1.059 $\pm$ 0.137 & 1.214 $\pm$ 0.130 & 1.143 $\pm$ 0.237 & 1.026 $\pm$ 0.207 & 0.816 $\pm$ 0.174 & 0.289 $\pm$ 0.178 & 0.624 $\pm$ 0.338 \\
C$_{1}$ & 1.190 $\pm$ 0.129 & 1.120 $\pm$ 0.134 & 1.329 $\pm$ 0.230 & 0.774 $\pm$ 0.195 & 1.066 $\pm$ 0.165 & 0.459 $\pm$ 0.183 & 0.441 $\pm$ 0.353 \\
C$_{2}$ & 1.222 $\pm$ 0.178 & 1.105 $\pm$ 0.188 & 1.339 $\pm$ 0.316 & 0.918 $\pm$ 0.274 & 1.137 $\pm$ 0.236 & 0.549 $\pm$ 0.234 & 0.553 $\pm$ 0.428 \\
D$_{1}$ & 1.078 $\pm$ 0.173 & 1.262 $\pm$ 0.189 & 1.646 $\pm$ 0.271 & 1.173 $\pm$ 0.232 & 0.940 $\pm$ 0.144 & 0.306 $\pm$ 0.174 & 0.909 $\pm$ 0.336 \\
D$_{2}$ & 1.050 $\pm$ 0.201 & 1.285 $\pm$ 0.228 & 1.771 $\pm$ 0.330 & 1.138 $\pm$ 0.295 & 0.908 $\pm$ 0.221 & 0.291 $\pm$ 0.228 & 1.032 $\pm$ 0.412 \\
\hline
\end{tabular}
\end{table*}

\begin{figure}
\includegraphics[width=\columnwidth]{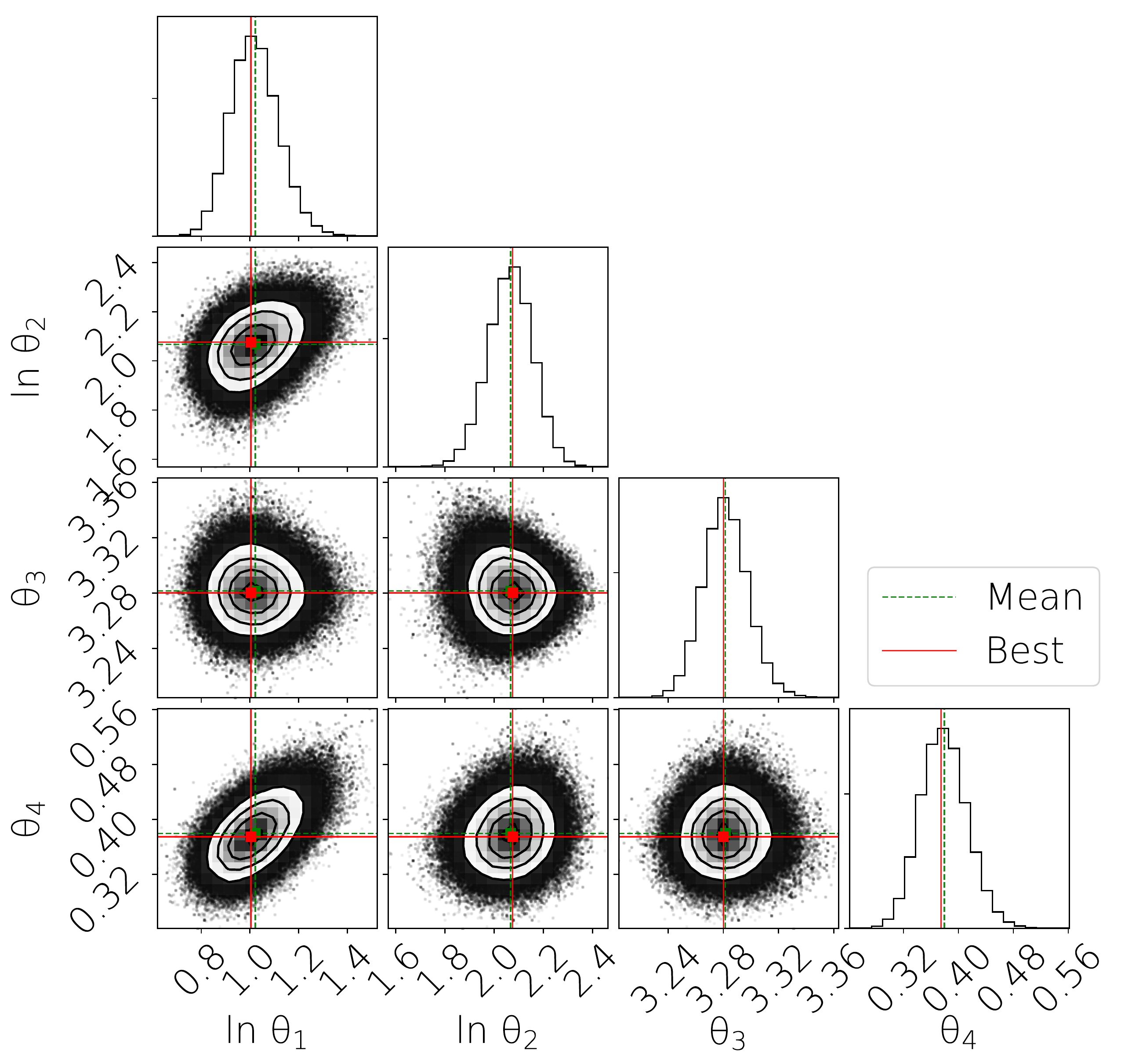}
\caption{Posterior distribution of the four GP hyperparameters returned by the MCMC sampling for a given dataset sampled with scheme A$_{1}$. The red solid lines indicate the best hyperparameter values, i.e. the values that maximize the likelihood of the MCMC. The green dashed lines indicate the median hyperparameter values over the posterior density. This plot was made using the \textsc{corner} python module \citep{Foreman-Mackey2016}.}
\label{fig:post_A1}
\end{figure}

\begin{figure}
\includegraphics[width=\columnwidth]{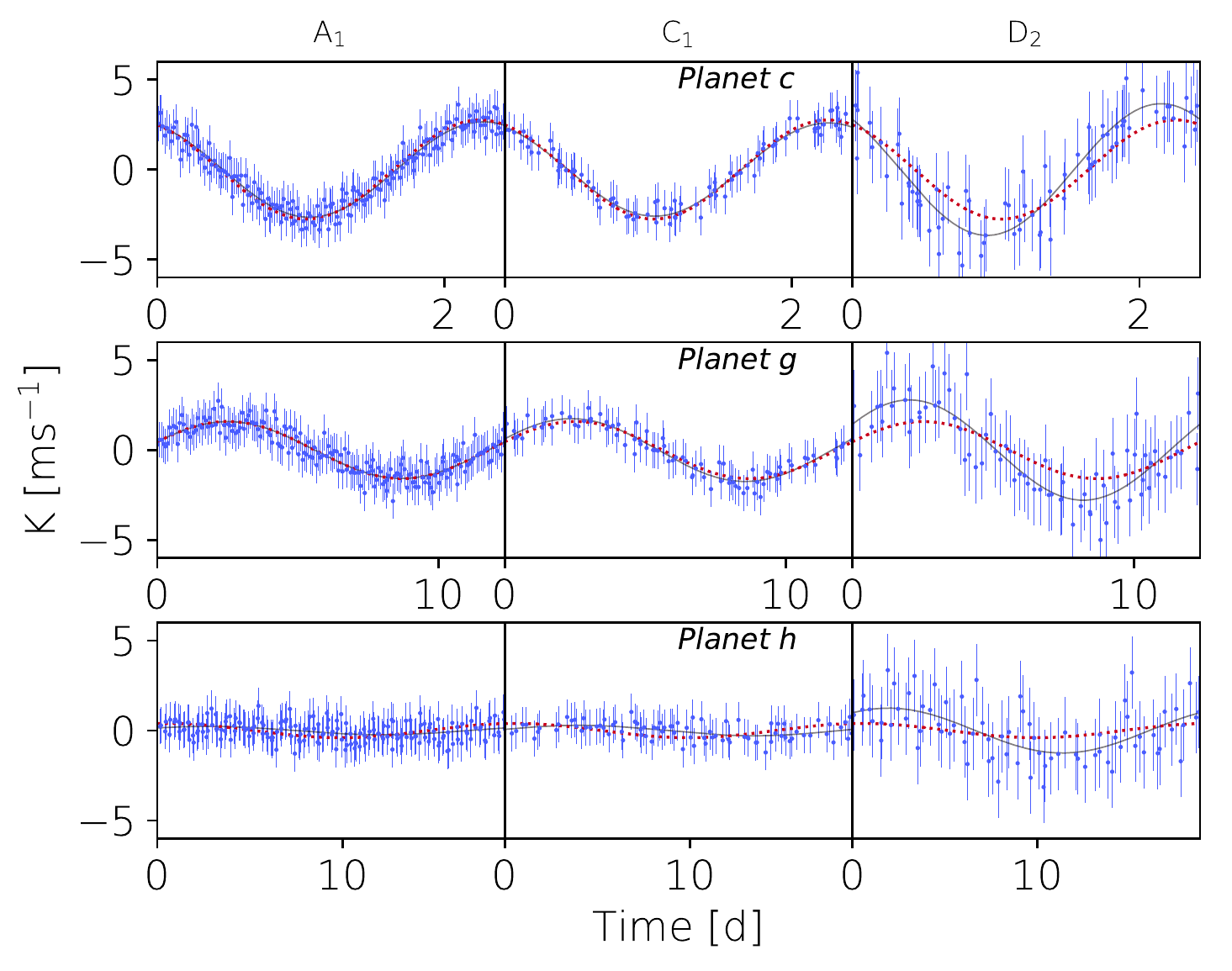}
\caption{Orbit-folded recovered RV curves for TRAPPIST-1 planets c (top row), g (middle row) and h (bottom row), and sampling schemes \A\ (first column), \C\ (middle column) and \Dd\ (right column) when stellar activity is taken into account, each for a given realization of the white noise. For each planet, the X-axis covers one orbital period on which all data points are folded.}
\label{fig:phase_fold_pl}
\end{figure}

\begin{figure}
\includegraphics[width=\columnwidth]{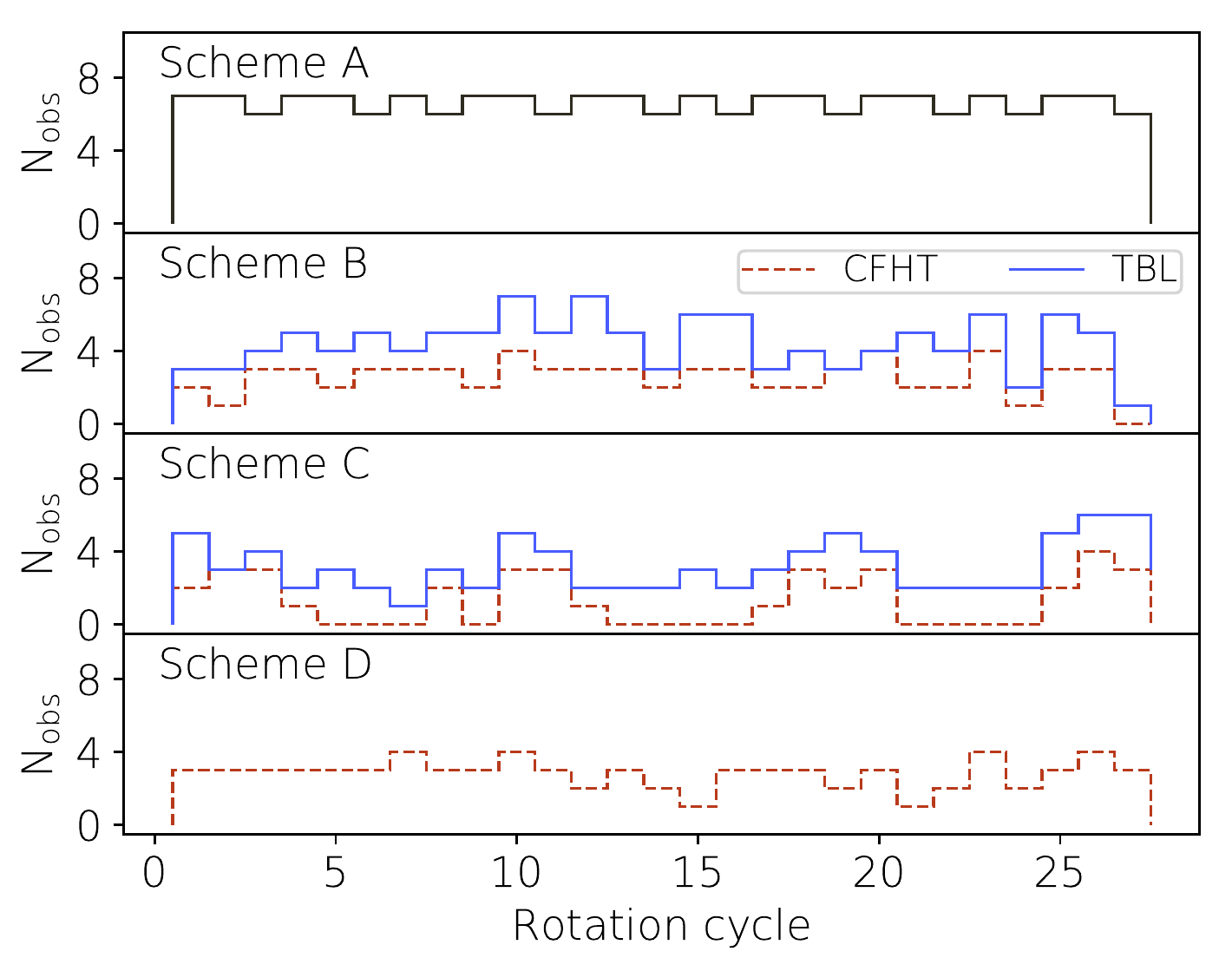}
\caption{Histograms of the number of observations (N$_{\mathrm{obs}}$) per stellar rotation cycle for each sampling scheme. When relevant, the observations are shown in dashed red and solid blue lines when from CFHT and TBL respectively.}
\label{fig:phase_fold_act}
\end{figure}


The fit to the RV time-series is noticeably degraded for sampling scheme D. In this case, the MCMC sampling of the parameter space does not converge anymore. The GP prefers to either minimize the smoothing factor $\theta_{4}$ or shrink the timescale $\theta_{2}$ so that the correlation between consecutive data points becomes ridiculously small leading to a non-physical modeling of the stellar activity RV signal. This implies that the RV curve is no longer sampled at a high enough rate to allow for a reliable fit of the data with the 4-hyperparameter GP we used.

When no stellar activity component is added to the data, datasets sampled on the \D\ scheme provide unbiased mass estimates with a precision of $>$5~$\sigma$ for all TRAPPIST-1 planets, demonstrating that sampling D performs a satisfactory coverage of the planetary phase curves. This is also notable in Fig~\ref{fig:phase_fold_pl} showing examples of orbit-folded RV curves of planet signals for sampling schemes \A, \C\ and \Dd, with activity taken into account and filtered out from the data as outlines in Section~\ref{sec:3.3}. Fig~\ref{fig:phase_fold_act} shows the histograms of the number of observations per stellar rotation cycle for each considered sampling scheme. Scheme D performs worst with only $\sim$3 data points per stellar rotation period. This is an insufficient coverage of the stellar rotation cycle before the spot pattern evolves significantly, leading to the non-convergence of the MCMC process. To address this issue, we have to freeze two of the 4 GP parameters, namely the timescale and smoothing parameters, $\theta_{2}$ and $\theta_{4}$, to their best estimates when fitting the stellar activity signal only (see Table~\ref{tab:results_actonly}).



\begin{figure*}
  \begin{center}
    \subfloat[1 \ms\ rms white noise]{
      \includegraphics[height=9cm,width=\columnwidth]{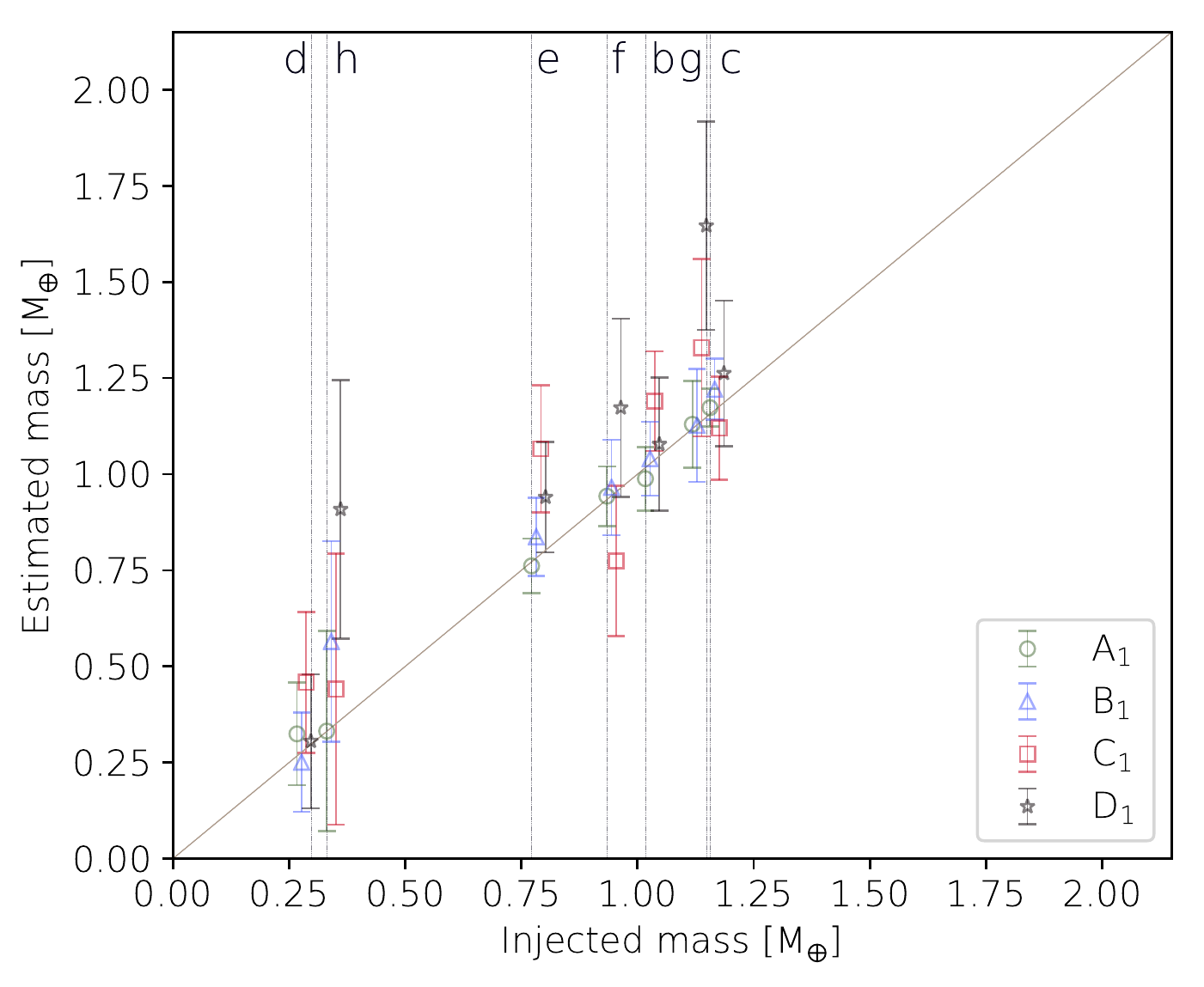}
      \label{sub:mass_mass_1}
                         }
    \subfloat[2 \ms\ rms white noise]{
      \includegraphics[height=9cm,width=\columnwidth]{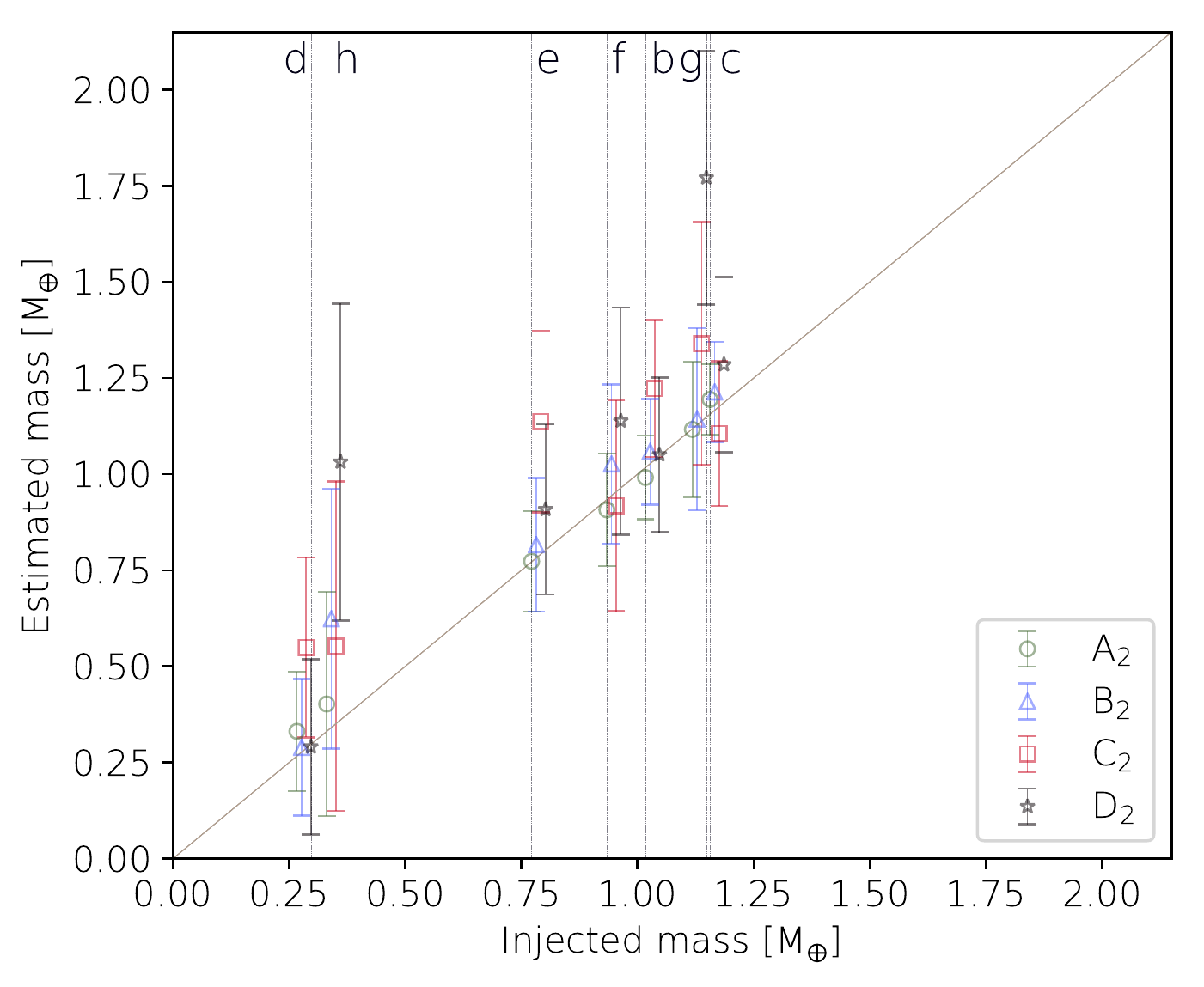}
      \label{sub:mass_mass_2}
                         }
    \caption{Estimated planetary masses as a function of the injected masses for the four sampling cases listed in Table \ref{tab:cases} at 1 \ms\ (left panel) and 2 \ms\ (right panel) rms white noise. For clarity purposes, we slightly shifted the mass estimates for the different sampling schemes to the left (for planets g and d) or to the right (remaining planets).}
    \label{fig:mass_mass}
  \end{center}
\end{figure*}

Fig \ref{fig:mass_mass} shows the mass estimates as a function of the injected masses for all the considered cases. With scheme D, the masses of planets g, f and h are strongly over-estimated, regardless of the white noise level. In contrast, the amplitude of the GP is lower than when fitting the stellar activity signal alone, implying an under-estimation of the latter (see Table \ref{tab:results_tot}). Moreover, the planet mass estimates are considerably less precise than those obtained from TTV except for planet b (see Table \ref{tab:masses}). In an attempt to improve the planet mass estimates while still using a single telescope, we doubled the number of observations per CFHT night. However, this configuration only marginally improves the precision on the mass estimates and the masses of planets g and h remain largely over-estimated. This demonstrates that doubling the number of observations per night in this configuration still leads to an incomplete coverage of the stellar rotation cycle. The same goes for sampling strategies aimed at phases where the planetary signal is expected to be maximum.


In contrast with scheme D, the fit of datasets sampled with scheme \B\ provides remarkably good mass estimates with unbiased estimates at a precision $>$8~$\sigma$ for planets b, c, g, f and e. However, the mass estimates of planets d and h are measured at a precision $<$3~$\sigma$ with an over-estimation of the mass of planet h. Case \Bb\ leads to unbiased mass estimates at a precision of $\ga$5~$\sigma$ for planets b, c, g, f and e. Compared to sampling B, scheme C provides mass estimates degraded in precision due to a significant decrease in the number of observations. However, the mass estimates for planets g and f remain much less over-estimated than in case D (see Fig~\ref{fig:phase_fold_pl}). For these planets in particular, the mass estimates are less degraded when moving from \C\ to \Cc\ than from \Cc\ to \Dd, which implies that the observational sampling mainly explains the erroneous mass estimates recovered in case \Dd. 

We finally generated $\sim$40 datasets sampled in case \C\ with different realisations of white noise and modelled each of them using the procedure described in Section~\ref{sec:3.3}, but this time freezing the planet orbital phases to their best estimates from \citetalias{delrez2018}. For this sampling scheme, it turns out that freezing the planet orbital phases when modelling the dataset has virtually no impact on the detectability of the planets and no more then marginally increases the precision on the planet masses (by about 2~\%).

\subsection{Planet detection}

\begin{figure*}
  \begin{center}
    \subfloat[1 \ms\ rms white noise]{
      \includegraphics[height=9cm,width=\columnwidth]{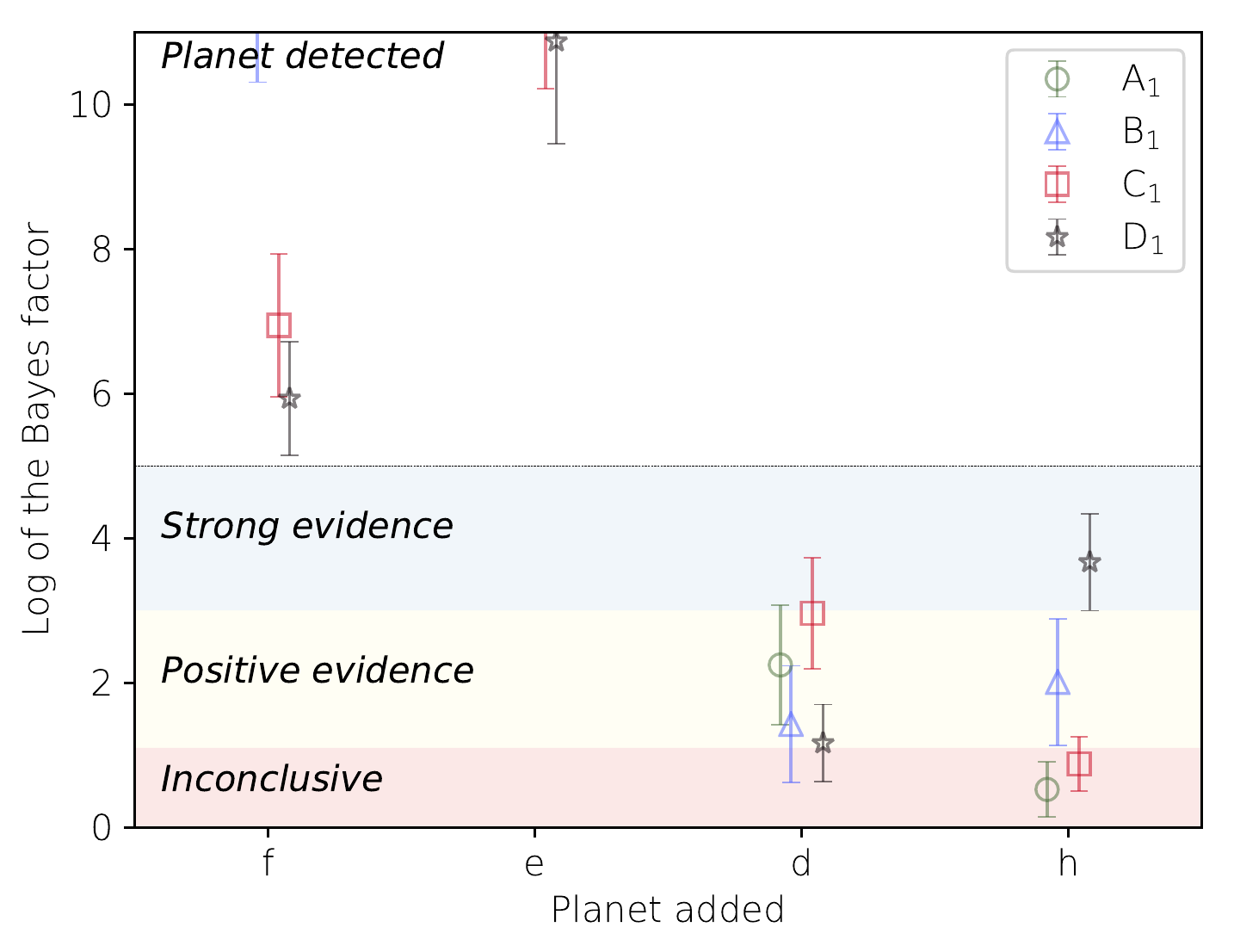}
      \label{sub:ML_1}
                         }
    \subfloat[2 \ms\ rms white noise]{
      \includegraphics[height=9cm,width=\columnwidth]{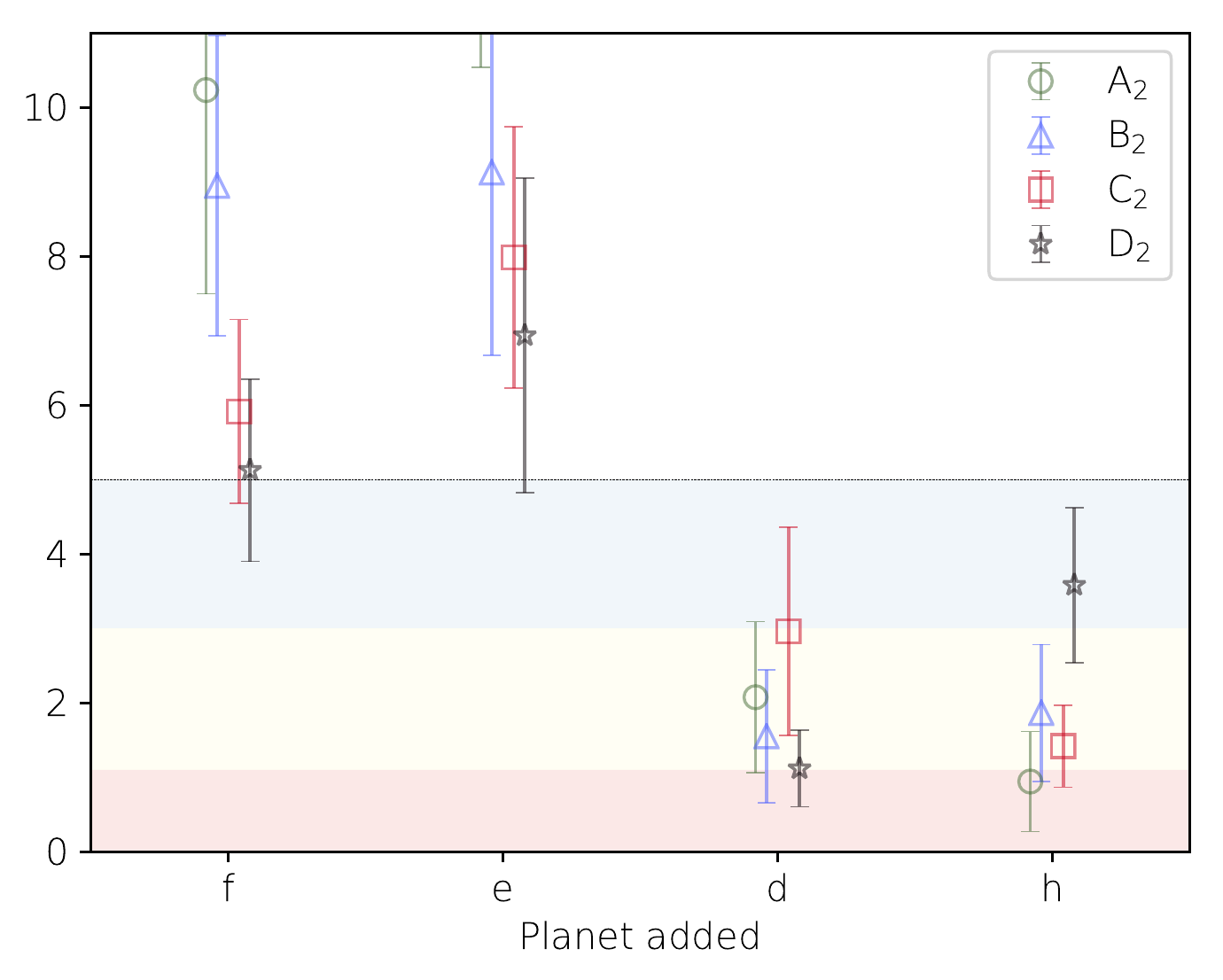}
      \label{sub:ML_2}
                         }
    \caption{Logarithm of the Bayes factors relative to planets f, e, d and h for the four sampling schemes listed in Table \ref{tab:cases} at 1 and 2 \ms\ rms white noise level (left and right panels respectively). The blue, yellow and red bands correspond to strong, positive and non-conclusive evidence according to \citet{Jeffreys1961} criterion. A Bayes factor that lies beyond the black horizontal line is interpreted as a fair detection. The values that do not appear in the figure lie beyond the Y-axis upper limit.}
    \label{fig:ML}
  \end{center}
\end{figure*}

We computed the Bayes factors in relative to each of the seven TRAPPIST-1 planets added in a decreasing order of RV semi-amplitude for datasets sampled with the schemes listed in Table~\ref{tab:cases} with white noise levels of 1 and 2~\ms\ rms. We ran our estimation algorithm on 50 datasets with different realisation of white noise and computed the average and standard deviation of the distribution using a 3-$\sigma$ clipping process to prevent the average and standard deviation from being affected by strong outliers.

As a result, planets b, c, g, f and e are well detected at a white noise level of 1~\ms. When the white noise level is increased to 2~\ms, we start to loose planet f with sampling scheme  D and, to a lesser extent, with sampling scheme C. In contrast, the evidence for planets d and h lie systematically below the detection threshold. Fig~\ref{fig:ML} shows the Bayes factors in favor of planets f, e, d and h depending on the sampling scheme and white noise level. For each sampling scheme, we note similar trends for the evolution of the Bayes factor as a function of the planet searched at white noise levels 1 and 2~\ms. For sampling scheme A (i.e. evenly sampled datasets), all planets are firmly detected except planets d (positive evidence) and h (inconclusive). The evidence for planet d decreases with the number of data points except for sampling C where the mass of planet d is over-estimated. Finally, the Bayes factor of planet h is significantly higher is scheme D than in other schemes. This trend is likely correlated to the over-estimation of the mass of planet h, whose accuracy strongly decreases from schemes A/B/C to D. This observation needs however to be nuanced as planet h remains undetected, in average, for all the sampling schemes.

\section{Discussions and conclusion}\label{sec:section5}

In this paper, we simulated RV follow-ups of TRAPPIST-1 using various sampling strategies spread on the same 90-d period. We first generated a RV curve for TRAPPIST-1a including a realistic stellar activity signal generated from a non-magnetic stellar activity model statistically compatible with the K2 light curve \citep{luger2017} and the planet RV signal using the parameters measured from photometric analyses \citep{delrez2018,grimm2018}. This resulted in planet and stellar activity curves with similar amplitudes ($\sim$5~\ms). We then built time-series using various sampling schemes, namely evenly sampled observations, bi-site observations from complementary longitudes and observations from a single location, to which we added a random noise accounting for additional noise sources (photon and instrumental non-correlated noise).

The stellar activity curve is modeled using GPR with a quasi-periodic kernel. Assuming that the planet orbital periods are known from transit photometry, we obtain that masses can be estimated at a precision $\ga$5$\sigma$ for planets b, c, g and e at a white noise level of 1~\ms, regardless of the considered sampling scheme. Planets d and h are systematically undetected. The increase in the white noise level impacts the errors on the planet masses in different ways among the error bars on the planet mass estimates, some being dominated by the white noise and only poorly-sensible to the stellar activity signal (planets c, f and e) while others are dominated by the stellar activity (mostly planets b, d and h).

Ancillary stellar activity indicators could be included to the GP modelling of the stellar activity signal, in a similar approach as \citet{rajpaul2015,jones2017}. However, the sensitivity of each indicator to stellar activity is shown to vary from one indicator to another, as demonstrated by the obvious differences observed between the most prominent peaks in the periodograms of RV and ancillary indicators time-series \citep{hebrard2016}. As a consequence, we ignored them when modelling the stellar activity signal.


This paper demonstrates that observational sampling is at least as critical as RV precision for velocimetric follow-ups of transiting planetary systems. For multi-planet TRAPPIST-1 analogs, we show that a dense-enough coverage of both planet orbital and stellar rotation cycles is essential to accurately recover the masses of all planets whose semi-amplitude RV wobble exceeds 1 \ms.  Failing to achieve so can cause some of the system planets to remain undetected, or lead to tentative planet detections with erroneous mass estimates.


In the specific case of TRAPPIST-1, we find that an uninterrupted  monitoring from CFHT can yield reliable mass estimates for planets with strongest RV semi-amplitudes and/or for which mass estimates are limited by the white noise (planets b and c, and to a lesser extent g, f and e).  However, planets with weaker RV signatures (d and h in particular, and potentially g and f as well) are likely to remain undetected, or marginally detected with erroneous mass estimates.  Limiting observations to bright-time periods only is expected to further degrade the situation, whereas adding data from a second observing site with similar instrumental capabilities and located at a complementary longitude can provide most of the missing material for reliably recovering some of the less massive planets. In this respect, SPIRou at CFHT and SPIP, its upcoming twin at TBL, are expected to form a powerful duo capable of achieving dense-enough coverage for the velocimetric follow-up of transiting planet candidates to be carried out within the SLS.

Besides that, we consider to adapt this study to other SLS-TF targets, especially young very active stars, in order to investigate how our results will be affected by significantly different activity and planet RV curves and prepare future observations with SPIRou.

\section*{Acknowledgements}

This project received funding from the European Research Council
(ERC) under the H2020 research \& innovation programme (grant
agreements \#740651 NewWorlds). All plots in this paper were created using the \textsc{matplotlib} python module \citep{hunter2007} and the ephemerides were computed using the \textsc{pyephem} module. We warmly thank the referee, Prof. Suzanne Aigrain, for valuable comments and suggestions which helped us reaching a clearer and more robust version of the manuscript. The authors finally thank Herv\'e Carfantan and M\'egane Boudineau for their help on the MCMC process.




\bibliographystyle{mnras}
\bibliography{ms.bbl}



\appendix
\section{Fit of the data}\label{app:A}

\begin{figure*}
\centering
\includegraphics[width=\linewidth]{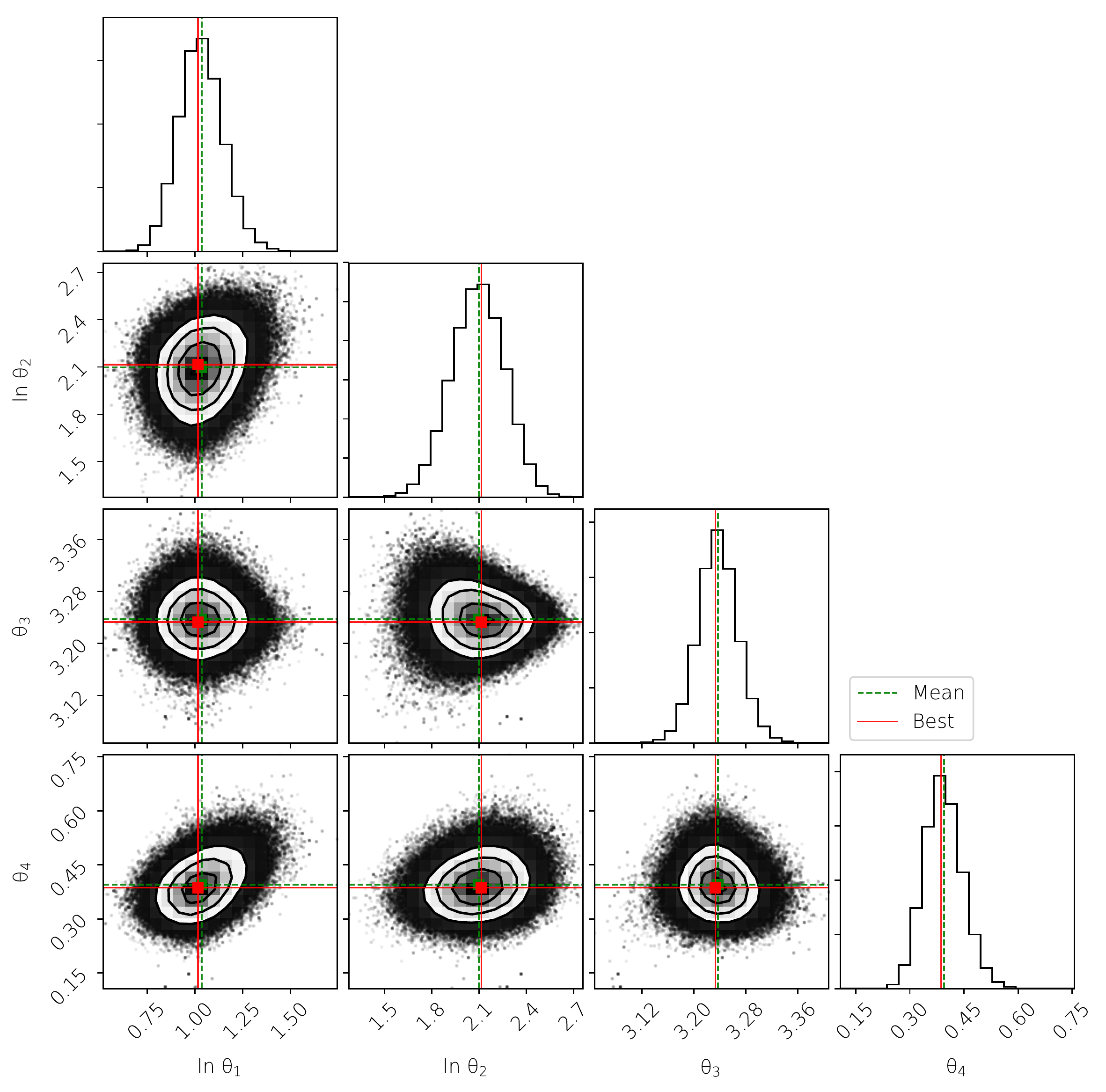}
\caption{Same as Fig \ref{fig:post_A1}, in case \Aa.}
\label{fig:post_A2}
\end{figure*}
\begin{figure*}
\centering
\includegraphics[width=\linewidth]{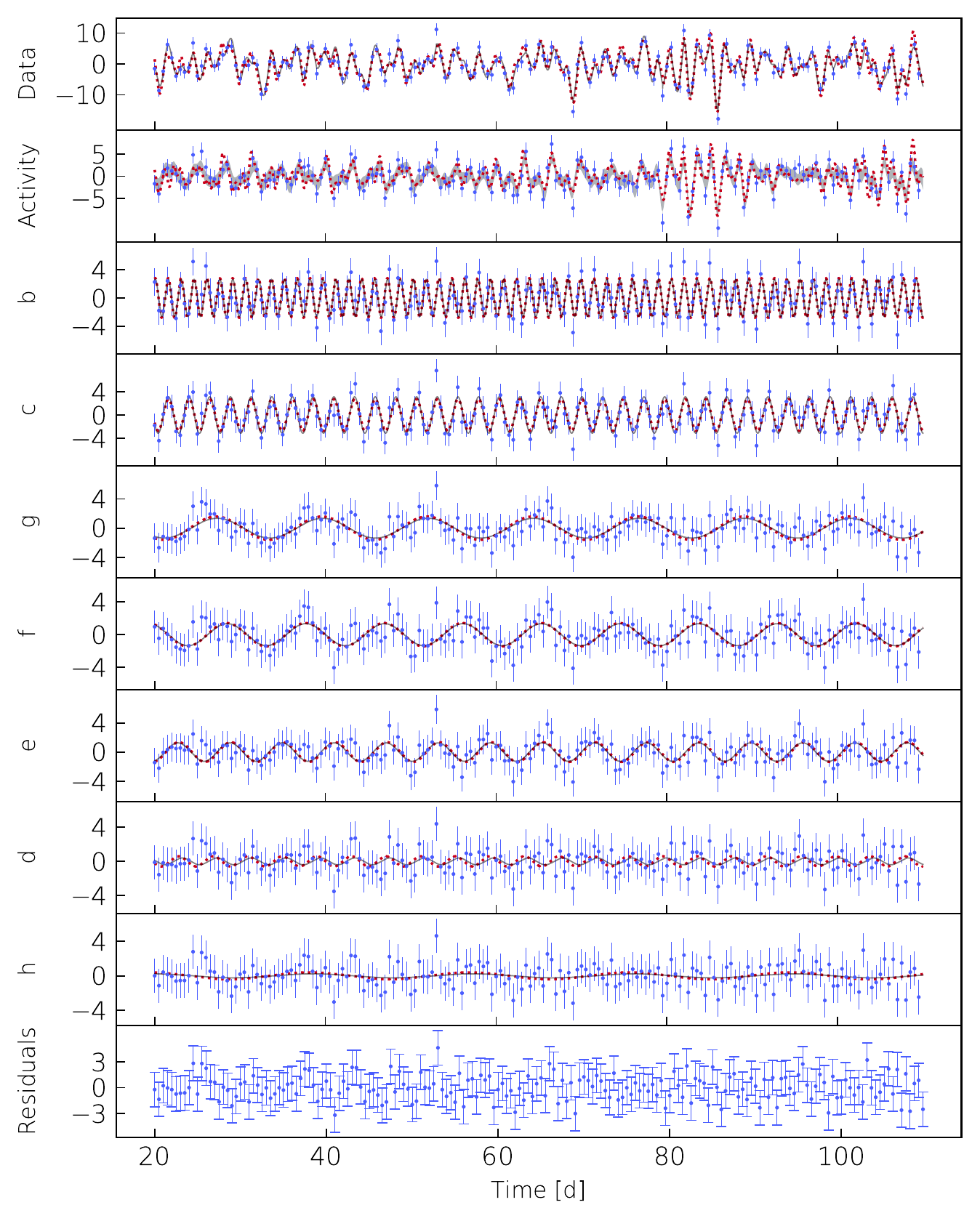}
\caption{Same as Fig \ref{fig:prediction_A1}, in case \Aa\ (rms of the residuals: 1.4 \ms).}
\label{fig:pred_A2}
\end{figure*}

\begin{figure*}
\centering
\includegraphics[width=\linewidth]{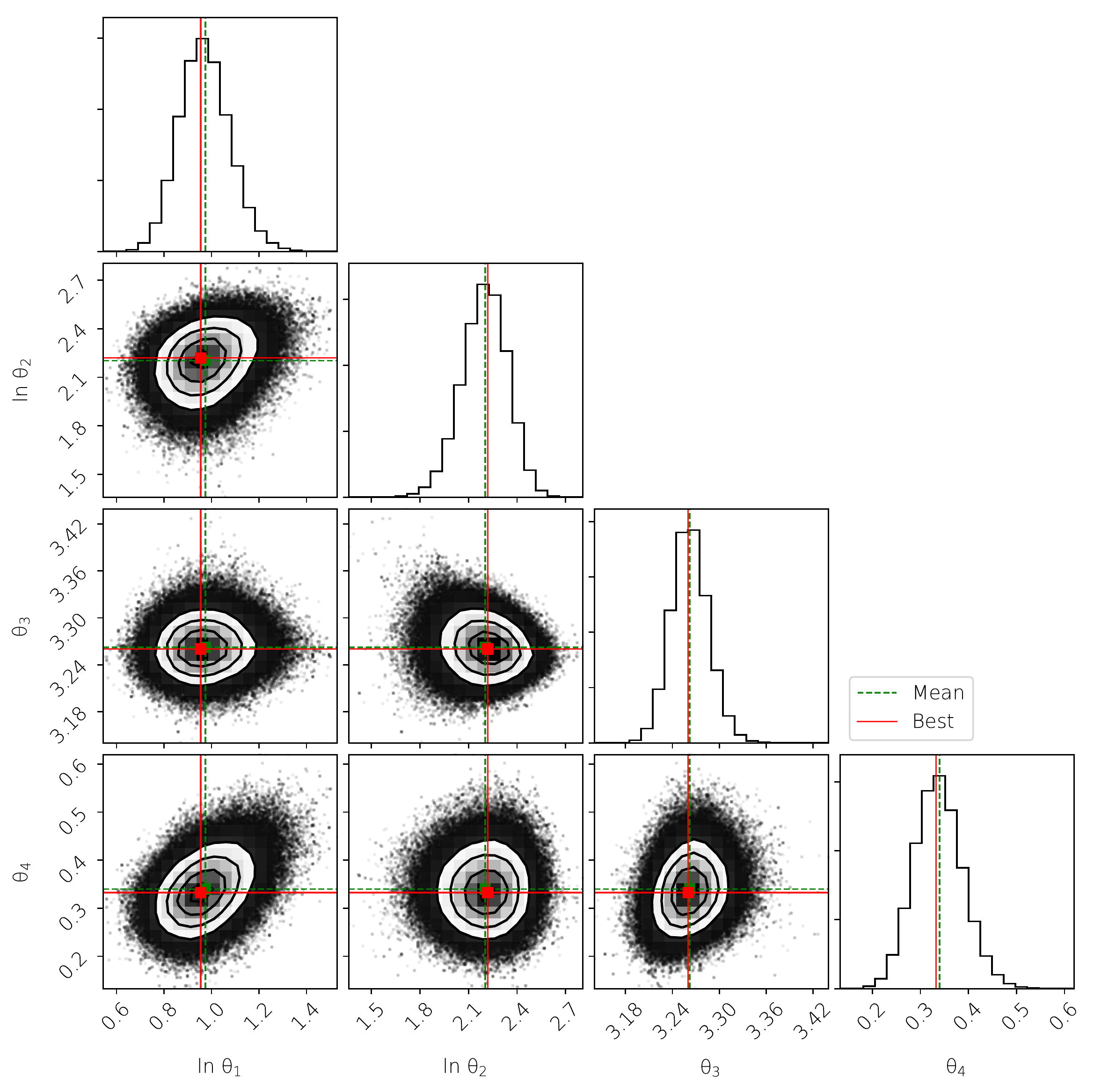}
\caption{Same as Fig \ref{fig:post_A1}, in case \B.}
\label{fig:post_B1}
\end{figure*}
\begin{figure*}
\centering
\includegraphics[width=\linewidth]{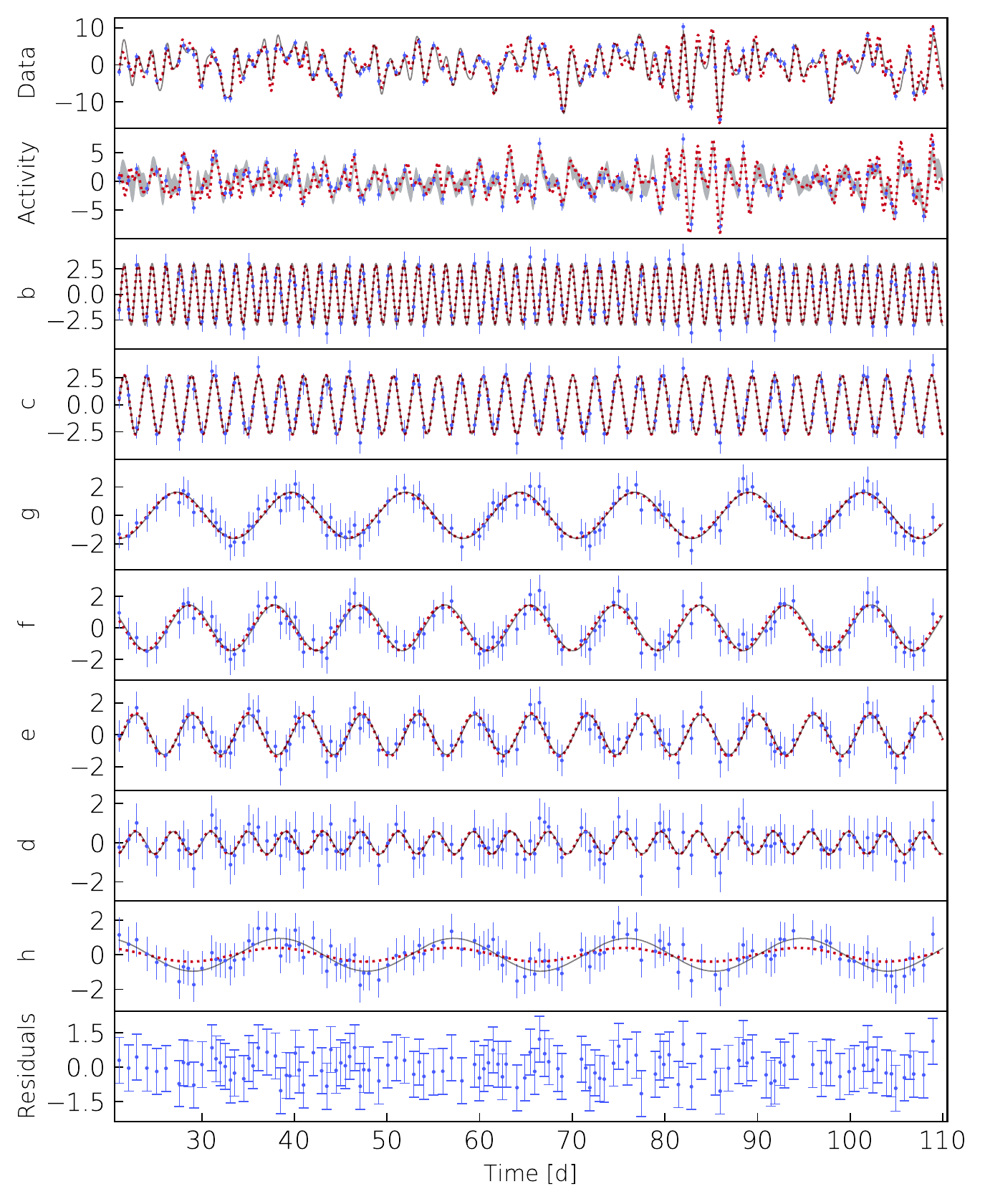}
\caption{Same as Fig \ref{fig:prediction_A1}, in case \B\ (rms of the residuals: 0.5 \ms).}
\label{fig:pred_B1}
\end{figure*}

\begin{figure*}
\centering
\includegraphics[width=\linewidth]{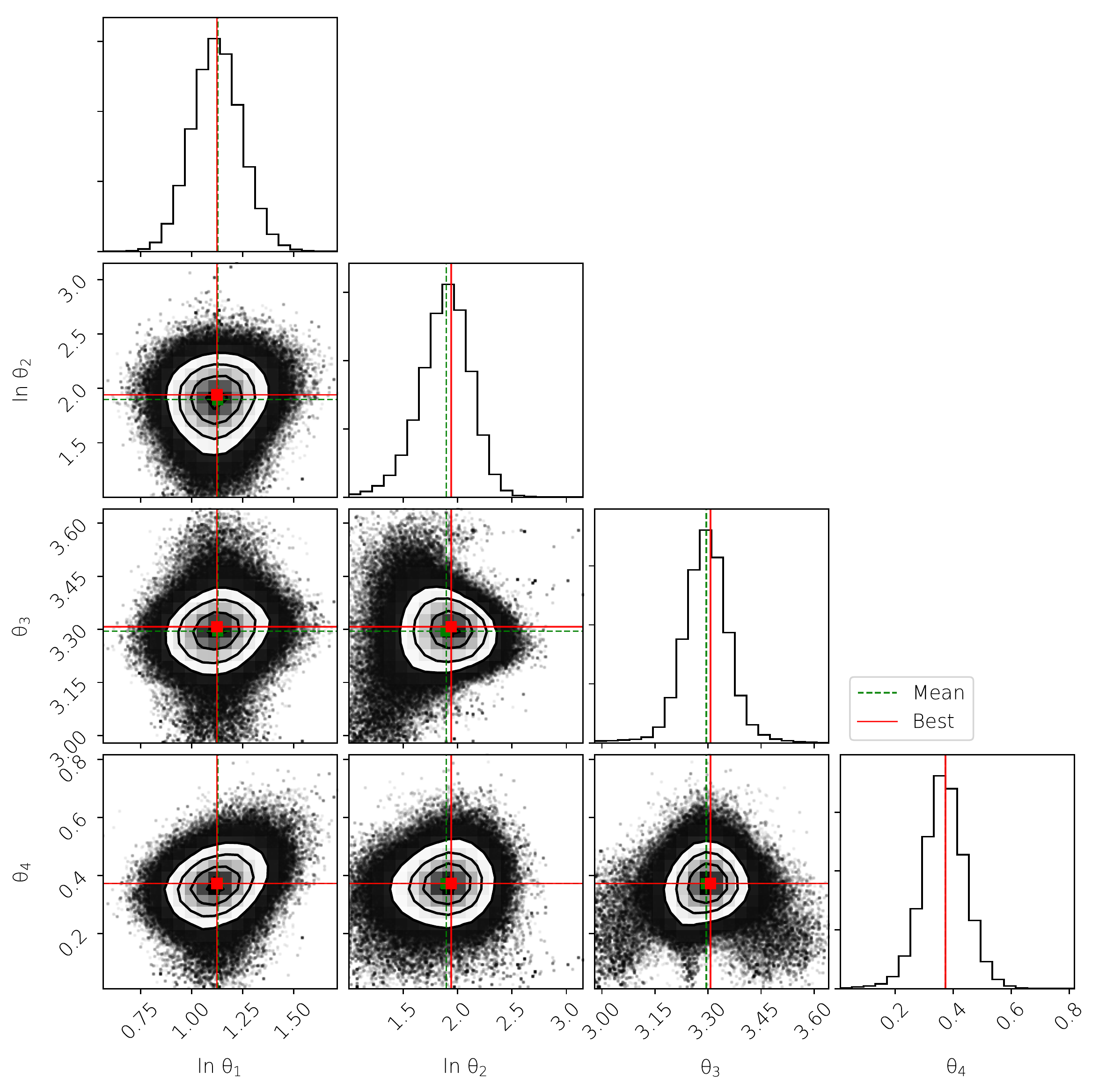}
\caption{Same as Fig \ref{fig:post_A1}, in case \Bb.}
\label{fig:post_B2}
\end{figure*}
\begin{figure*}
\centering
\includegraphics[width=\linewidth]{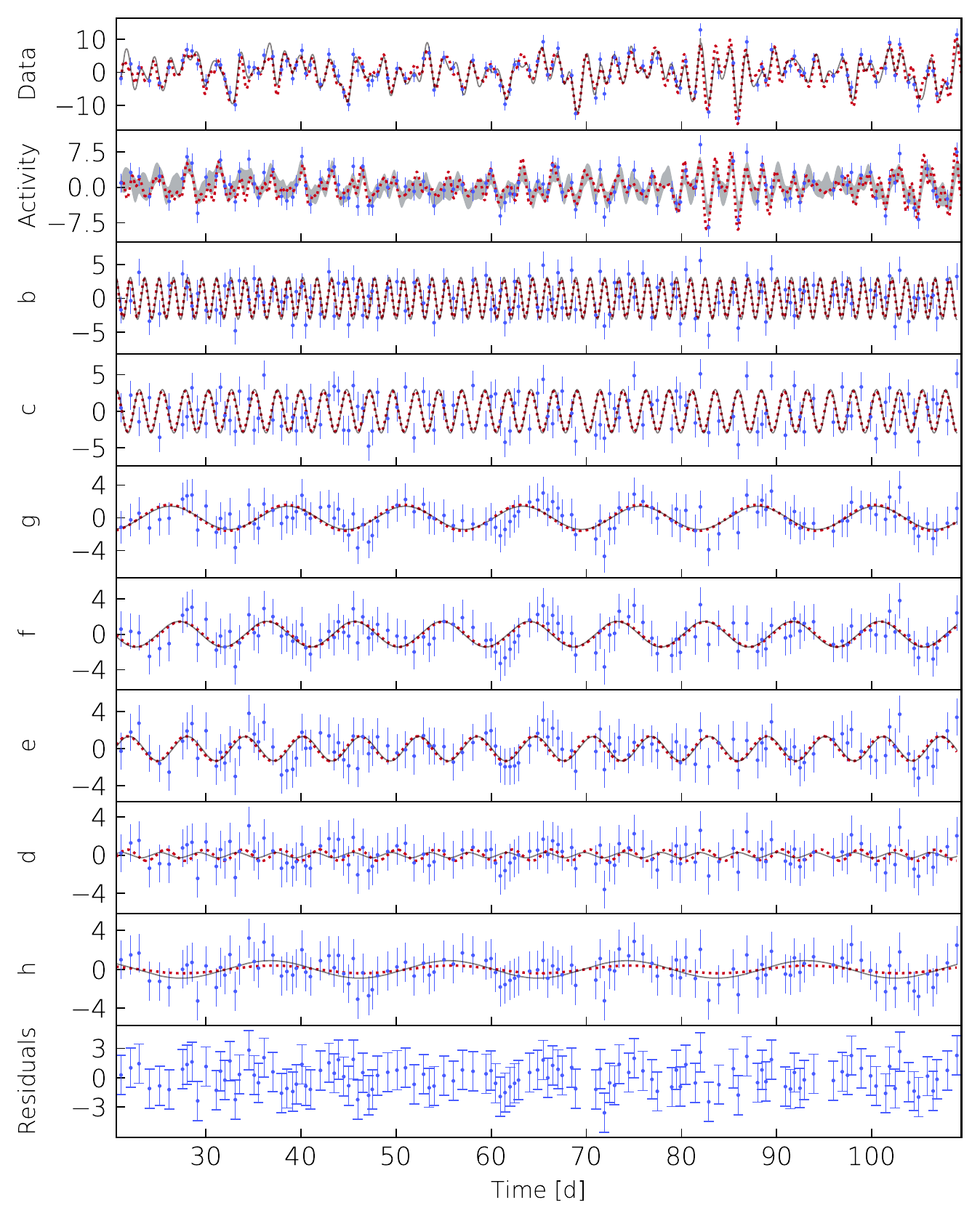}
\caption{Same as Fig \ref{fig:prediction_A1}, in case \Bb\ (rms of the residuals: 1.3 \ms).}
\label{fig:pred_B2}
\end{figure*}

\begin{figure*}
\centering
\includegraphics[width=\linewidth]{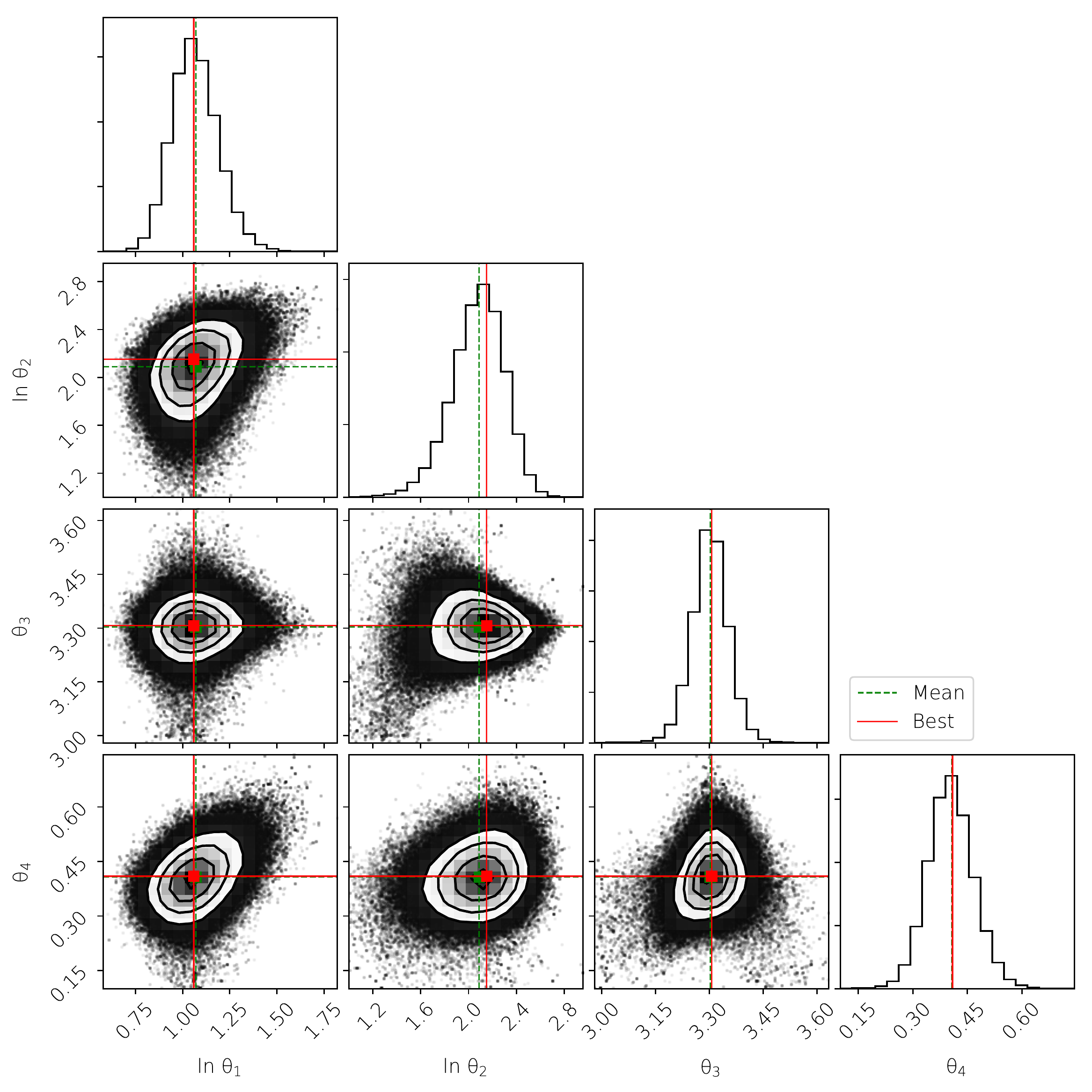}
\caption{Same as Fig \ref{fig:post_A1}, in case \C.}
\label{fig:post_Moon1}
\end{figure*}
\begin{figure*}
\centering
\includegraphics[width=\linewidth]{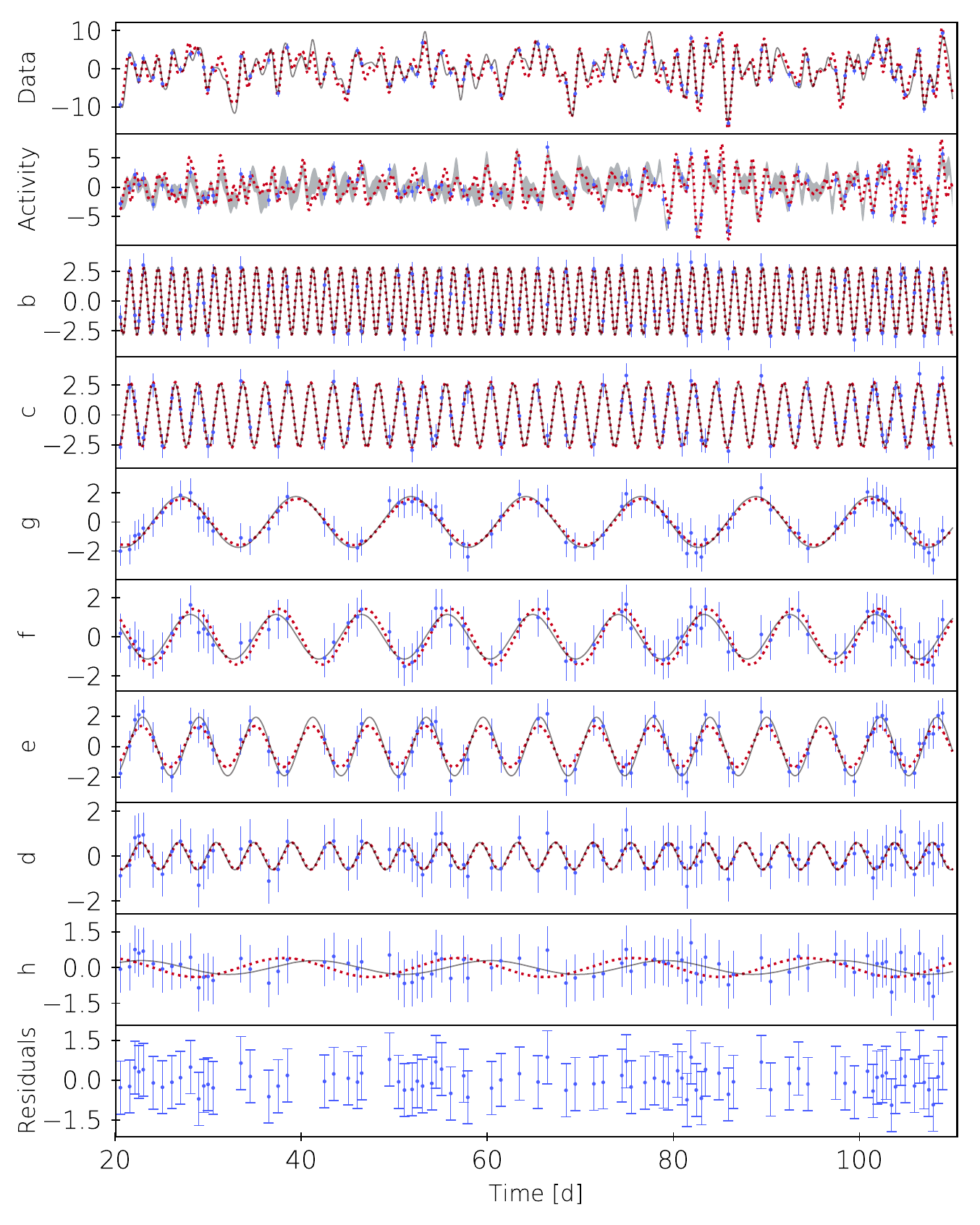}
\caption{Same as Fig \ref{fig:prediction_A1}, in case \C\ (rms of the residuals: 0.4 \ms).}
\label{fig:pred_Moon1}
\end{figure*}

\begin{figure*}
\centering
\includegraphics[width=\linewidth]{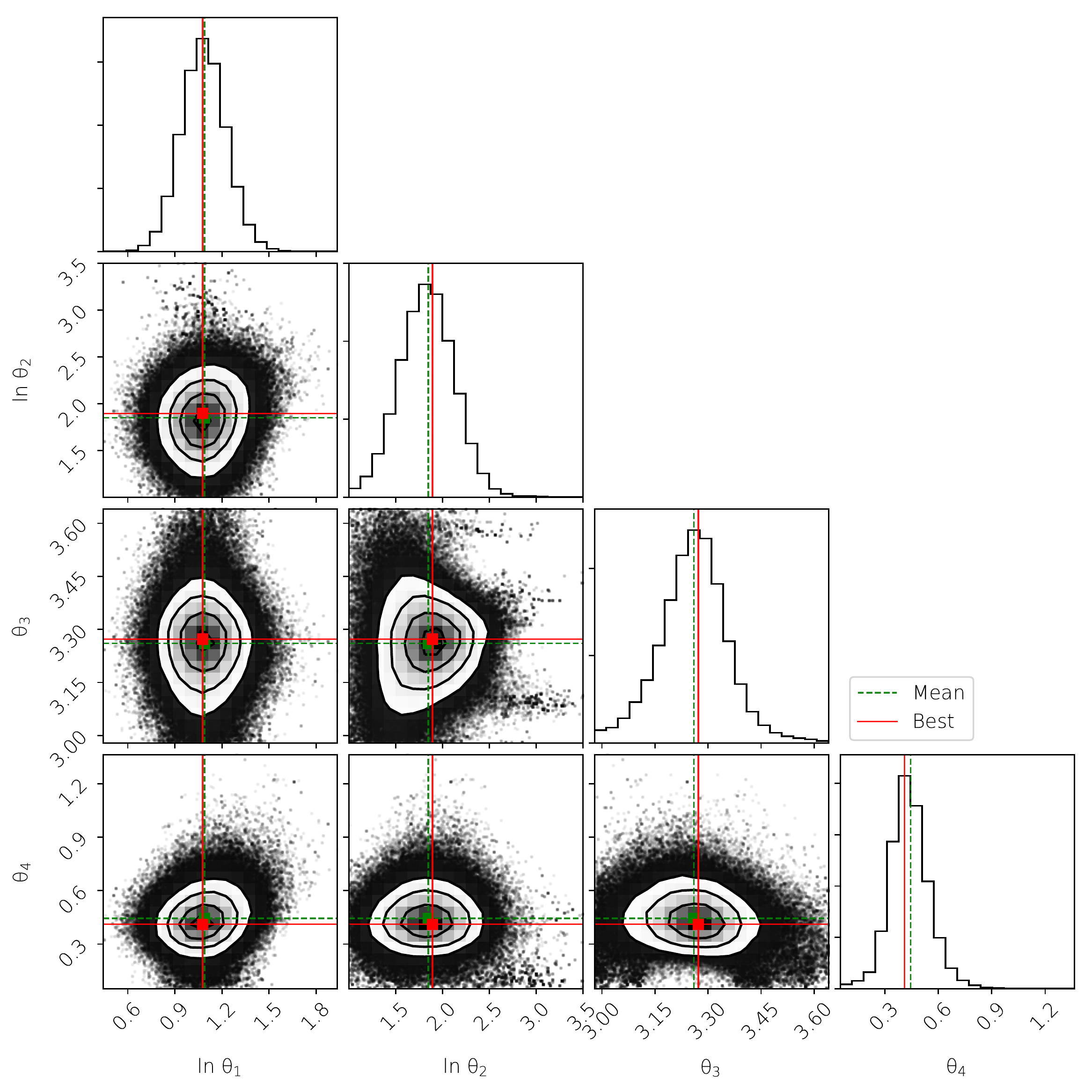}
\caption{Same as Fig \ref{fig:post_A1}, in case \Cc.}
\label{fig:post_Moon2}
\end{figure*}
\begin{figure*}
\centering
\includegraphics[width=\linewidth]{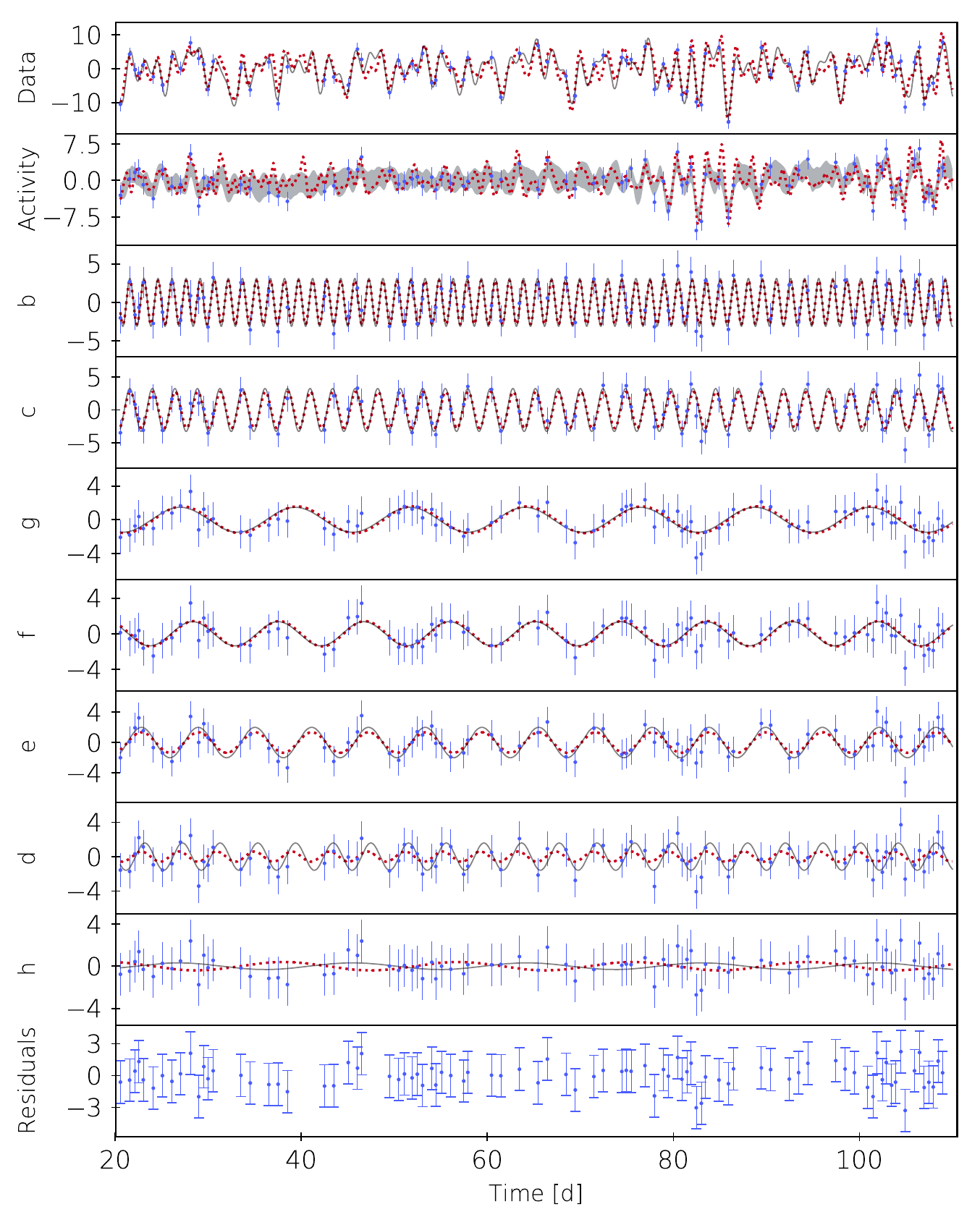}
\caption{Same as Fig \ref{fig:prediction_A1}, in case \Cc\ (rms of the residuals: 1.1 \ms).}
\label{fig:pred_Moon2}
\end{figure*}

\begin{figure*}
\centering
\includegraphics[width=\linewidth]{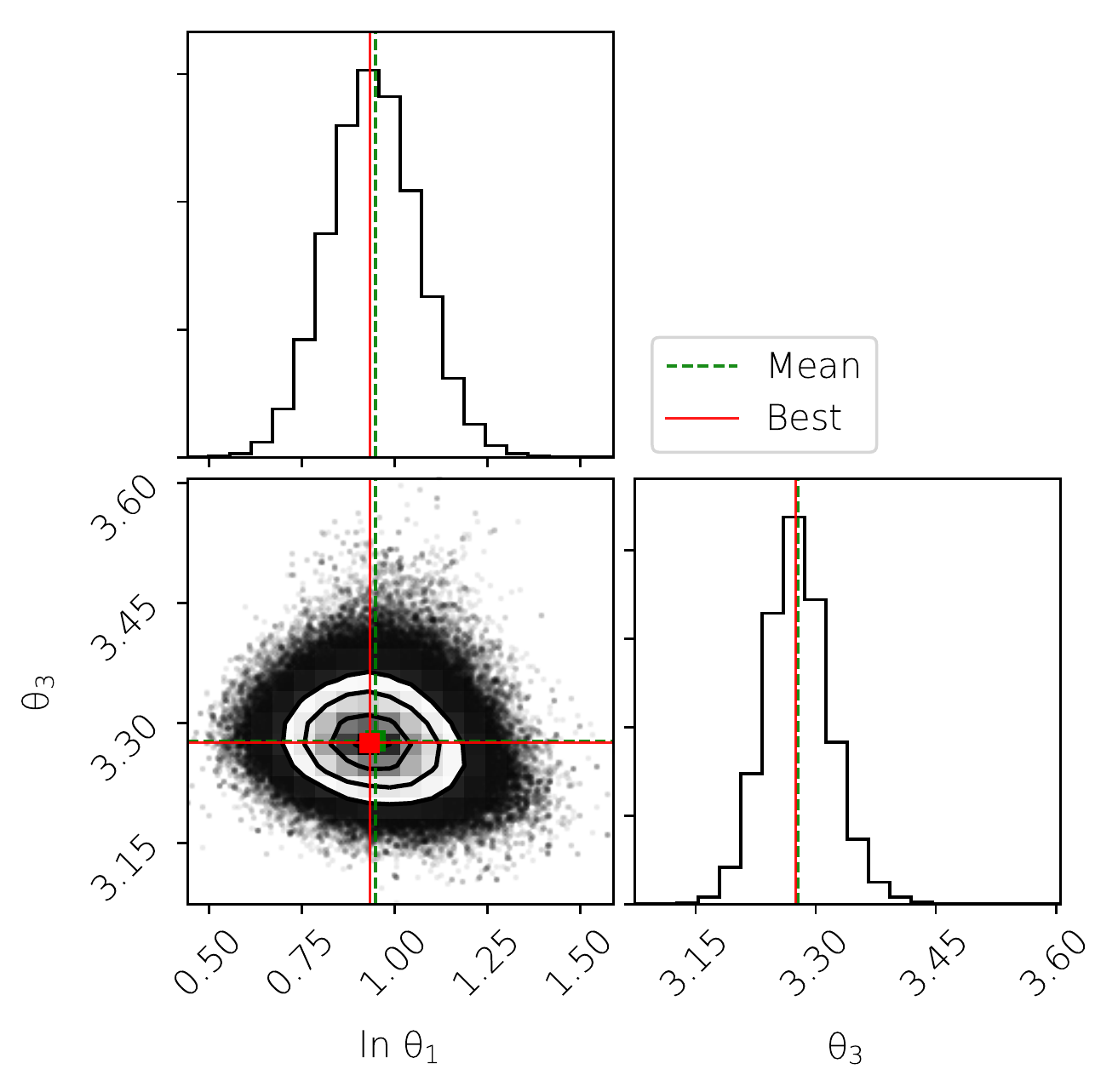}
\caption{Same as Fig \ref{fig:post_A1}, in case \D.}
\label{fig:post_C1}
\end{figure*}
\begin{figure*}
\centering
\includegraphics[width=\linewidth]{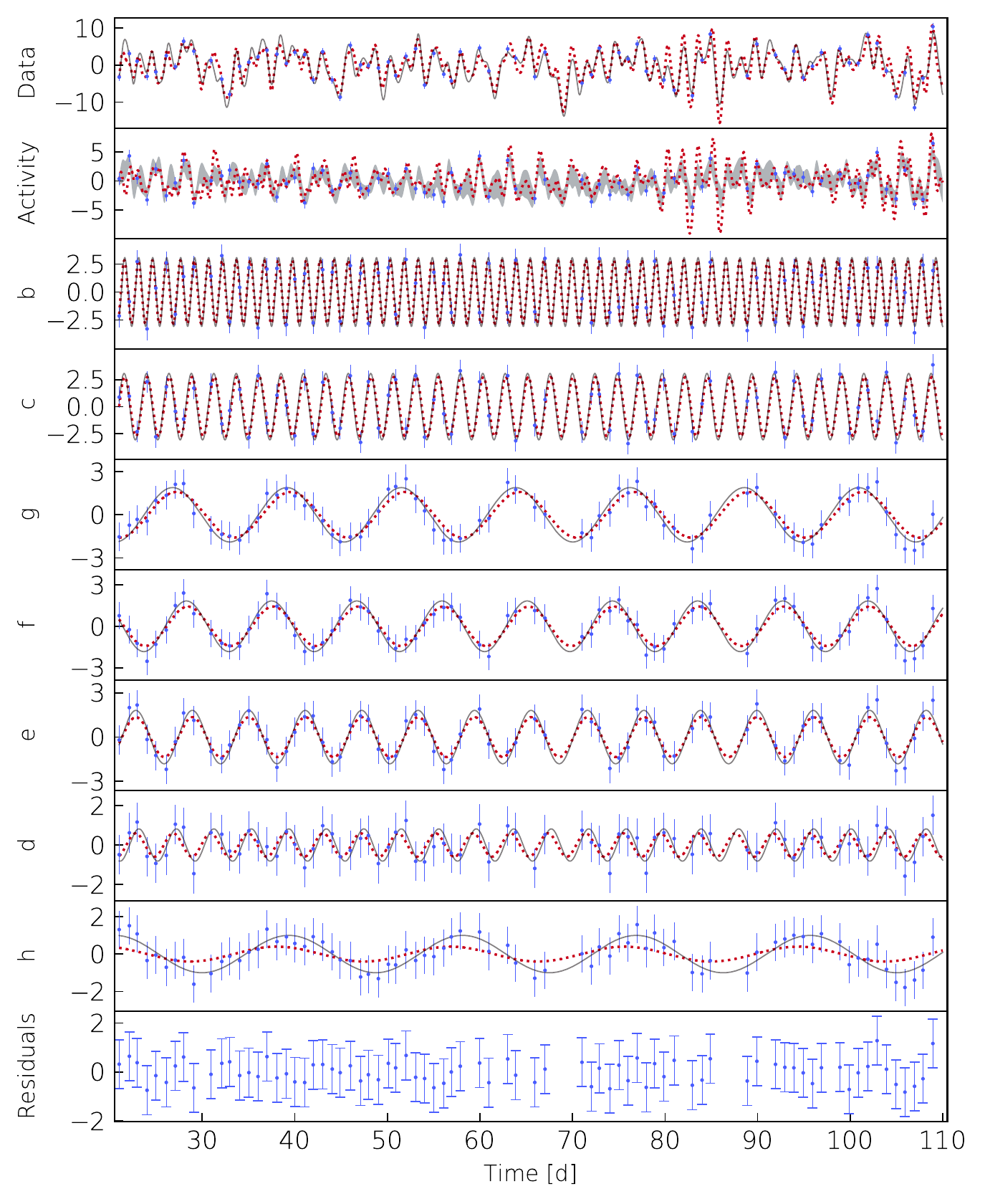}
\caption{Same as Fig \ref{fig:prediction_A1}, in case \D\ (rms of the residuals: 0.4 \ms).}
\label{fig:pred_C1}
\end{figure*}

\begin{figure*}
\centering
\includegraphics[width=\linewidth]{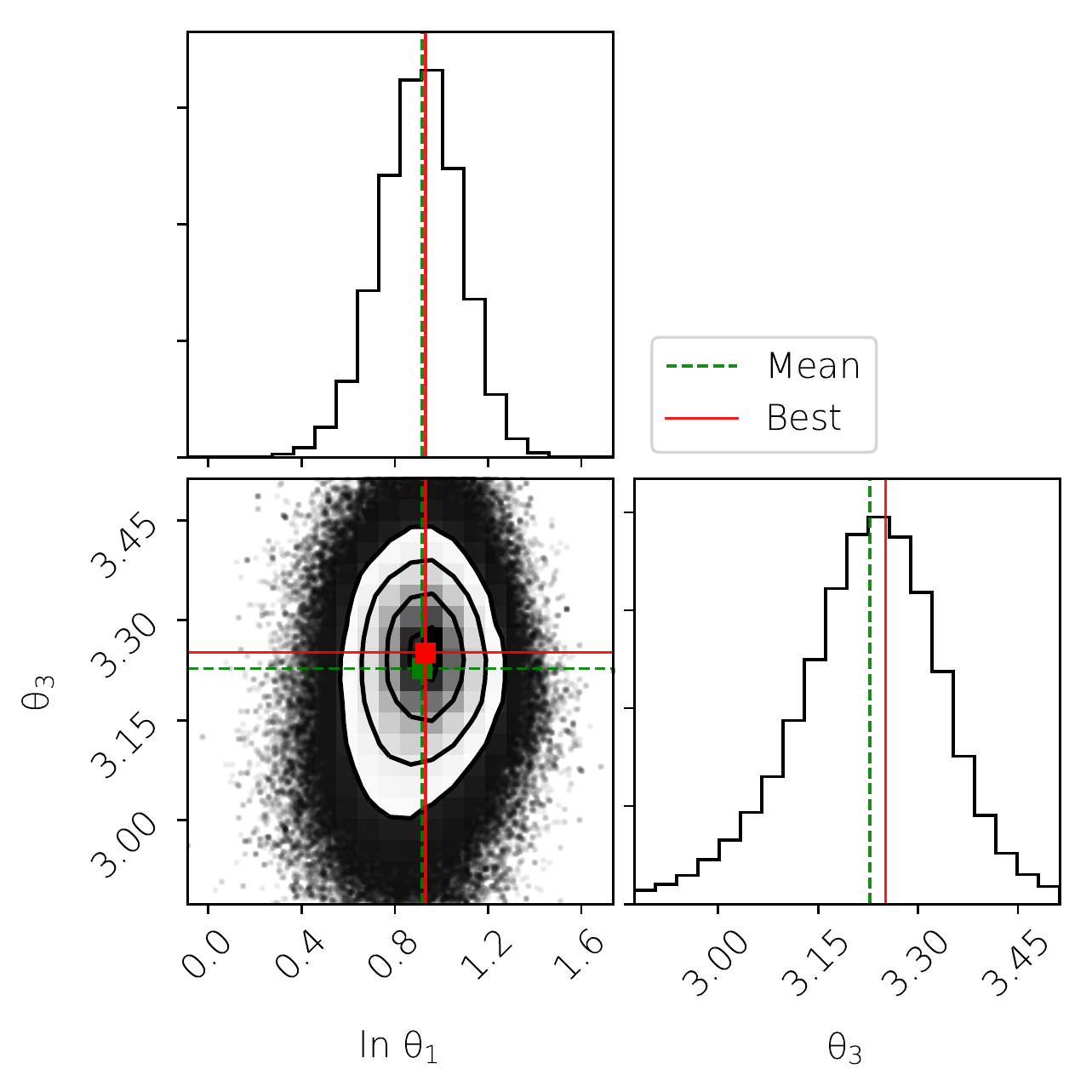}
\caption{Same as Fig \ref{fig:post_A1}, in case \Dd.}
\label{fig:post_C2}
\end{figure*}
\begin{figure*}
\centering
\includegraphics[width=\linewidth]{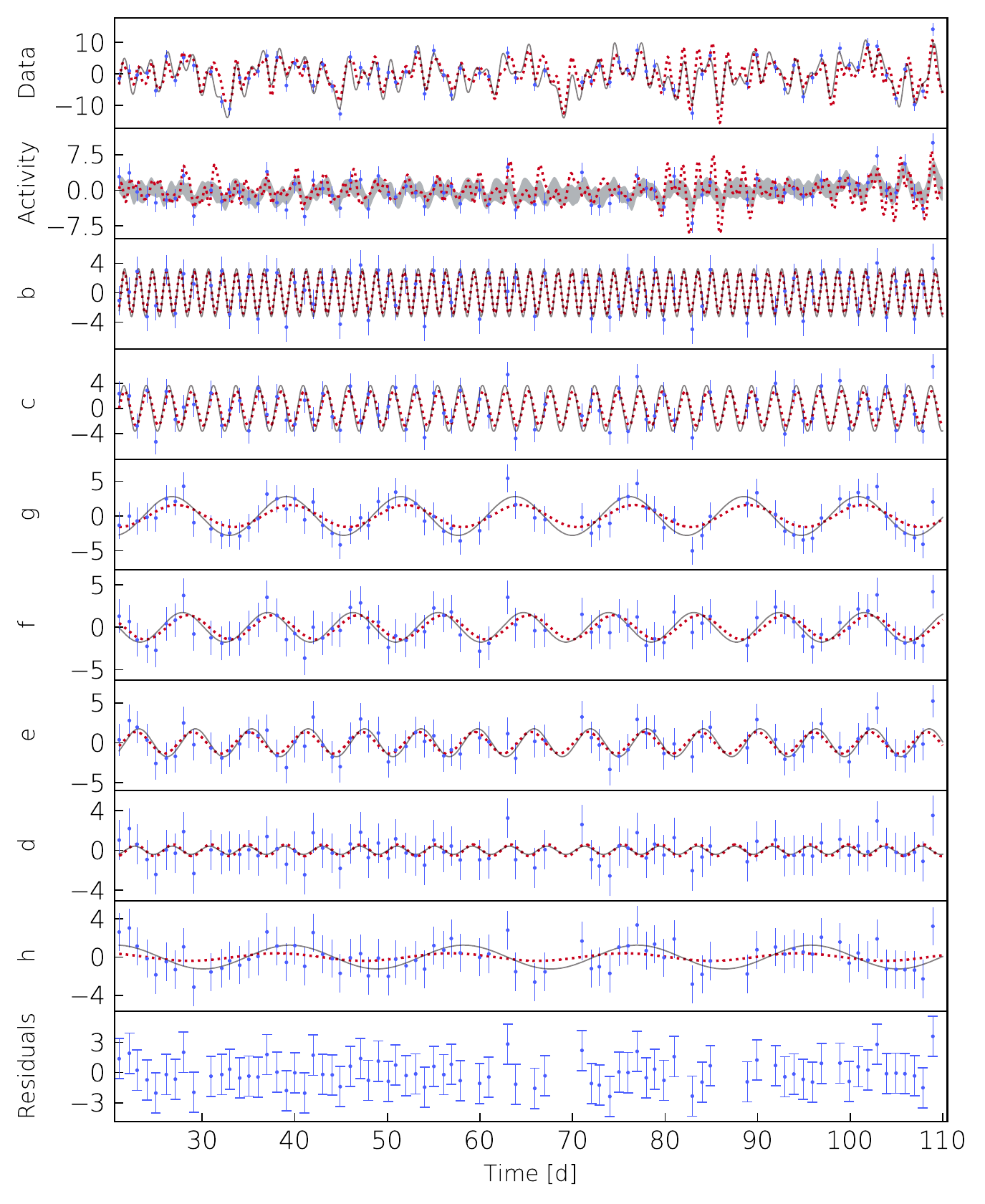}
\caption{Same as Fig \ref{fig:prediction_A1}, in case \Dd\ (rms of the residuals: 1.2 \ms).}
\label{fig:pred_C2}
\end{figure*}


\bsp	
\label{lastpage}
\end{document}